\documentclass{aa}  
\usepackage{graphicx}
\usepackage{xcolor}
\usepackage{natbib}
\usepackage{txfonts}
\usepackage{siunitx}
\usepackage{wasysym}
\usepackage{amsmath,amssymb}
\usepackage{booktabs}
\usepackage{longtable}
\newcommand{\abs}[1]{\ensuremath{\left\vert#1\right\vert}}
\newcommand{\vect}[1]{\vec{#1}}
\newcommand{\tensor}[1]{\tens{#1}}
\newcommand{\dif}{\mathop{}\mathopen{}\mathrm{d}}
\newcommand{\bnabla}{\boldsymbol{\nabla}}
\newcommand{\grad}{\bnabla}
\newcommand{\ez}{\vect{e}_z}
\newcommand{\er}{\ensuremath{\vect{e}_r}}

\newcommand{\ri}{\ensuremath{r_\text{i}}}
\newcommand{\ro}{\ensuremath{r_\text{o}}}
\newcommand{\rmix}{\ensuremath{r_\text{mix}}}
\newcommand{\rol}{\ensuremath{Ro_\ell}}
\newcommand{\roc}{\ensuremath{Ro_\text{c}}}
\newcommand{\npol}{\ensuremath{n}}
\newcommand{\Nrho}{\ensuremath{N_\varrho}}

\newcommand{\Prandtl}{\ensuremath{Pr}} 
\newcommand{\aspectratio}{\ensuremath{\chi}}
\newcommand{\Nu}{\ensuremath{Nu}}
 
\newcommand{\Ekman}{\ensuremath{E}}

\newcommand{\Di}{\ensuremath{\tilde{D}}}
\newcommand{\Ra}{\ensuremath{{Ra}}}
\newcommand{\RaM}{\ensuremath{{Ra^\textsc{m}}}}
\newcommand{\Rac}{\ensuremath{{Ra_\text{c}}}}
\newcommand{\Scond}{\ensuremath{S_{\!c}}}
\newcommand{\mrho}{\overline{\varrho}}
\newcommand{\mT}{\overline{T}}
\newcommand{\mP}{\overline{P}}
\newcommand{\parody}{\textsc{Parody}}
\newcommand{\magic}{\textsc{Magic}}

\newlength{\mafig}
\begin{document}

   \title{Gravity darkening in late-type stars. The Coriolis effect.}

   \subtitle{}

   \author{R. Raynaud\inst{1} 
     \and
     M. Rieutord\inst{2,3} 
     \and 
     L. Petitdemange\inst{4}
     \and 
     T. Gastine\inst{5}
     \and
     B. Putigny\inst{2,3}
   } 
   \institute{
     School of Astronomy, Institute for Research in
     Fundamental Sciences (IPM), P.O. Box 19395-5531, Tehran, Iran\label{i1}
     \and 
     Universit\'e de Toulouse; UPS-OMP; IRAP; Toulouse, France
     \and 
     CNRS; IRAP; 14, avenue \'{E}douard Belin, F-31400 Toulouse, France
     \and
     LERMA, Observatoire de Paris\thanks{R. Raynaud thanks the
       Observatoire de Paris for the granted access to the HPC
       resources of MesoPSL.}, PSL Research University, CNRS, Sorbonne
     Universit\'{e}s, UPMC Univ. Paris 06, \'{E}cole normale
     sup\'{e}rieure, F-75005, Paris, France\label{i3}
     \and
     Institut de Physique du Globe de Paris, Sorbonne Paris Cité,
     Université Paris-Diderot, UMR 7154 CNRS, 1 rue Jussieu, F-75005 Paris,
     France, \\ 
     \email{raphael.raynaud@ipm.ir, Michel.Rieutord@irap.omp.eu,
       ludovic@lra.ens.fr} }

   \date{Received ; accepted}

 
   \abstract 
   {Recent interferometric data have been used to constrain the
     brightness distribution at the surface of nearby stars, in
     particular the so-called gravity darkening that makes fast
     rotating stars brighter at their poles than at their
     equator. However, good models of gravity darkening are missing
     for stars that posses a convective envelope.}
   %
   {In order to better understand how rotation affects the heat
     transfer in stellar convective envelopes, we focus on the heat flux
     distribution in latitude at the outer surface of numerical
     models.}
   %
   {We carry out a systematic parameter study of three-dimensional,
     direct numerical simulations of anelastic convection in rotating
     spherical shells. As a first step, we neglect the centrifugal
     acceleration and retain only the Coriolis force. The fluid
     instability is driven by a fixed entropy drop between the inner
     and outer boundaries where stress-free boundary conditions are
     applied for the velocity field. Restricting our investigations to
     hydrodynamical models with a thermal Prandtl number fixed to
     unity, we consider both thick and thin (solar-like) shells, and
     vary the stratification over three orders of magnitude. We
     measure the heat transfer efficiency in terms of the Nusselt
     number, defined as the output luminosity normalised by the
     conductive state luminosity.}
   %
   {We report diverse Nusselt number profiles in latitude, ranging
     from brighter (usually at the onset of convection) to darker
     equator and uniform profiles. We find that the variations of the
     surface brightness are mainly controlled by the surface value of
     the local Rossby number: when the Coriolis force dominates the
     dynamics, the heat flux is weakened in the equatorial region by
     the zonal wind and enhanced at the poles by convective motions
     inside the tangent cylinder. In the presence of a strong
     background density stratification however, as expected in real
     stars, the increase of the local Rossby number in the outer
     layers leads to uniformisation of the surface heat flux
     distribution.}
   {}


   \maketitle
%

\section{Introduction}

Fifty years ago, \cite{lucy67} published a work on ``Gravity darkening
for stars with a convective envelope''. At the time, the motivation was
the interpretation of the light curves of the W Ursa Majoris
stars. Gravity darkening is indeed one of the phenomena that can
modify the surface brightness of a star and thus be important in the
interpretation of stellar light curves. Usually this phenomenon is
associated with fast rotating early-type stars. We recall that for
such stars, endowed with a radiative envelope, the flux varies with
latitude basically because their centrifugal flattening makes the
equatorial radius larger than the polar one. The temperature drop
between the centre and the pole or the equator of the star being
roughly the same, the temperature gradient is slightly weaker in the
equatorial plane. Hence, the local surface flux is slightly less at
the equator than at the poles; the equator appears darker
\cite[e.g.][]{monnier_etal07}. For many decades this phenomenon was
approximated by the \citet{zeipel1924} law:  $T_{\rm
  eff}\propto g_{\rm eff}^{1/4}$.  Sometimes, fitting data requires a
more general relation and von Zeipel's law was changed to $T_{\rm
  eff}\propto g_{\rm eff}^\beta$, and $\beta$ adjusted.

Observational works that have put constraints on the gravity-darkening
exponent $\beta$ come essentially from the photometry of eclipsing
binaries \cite[][]{dju06} and interferometric observations of fast
rotating stars \cite[e.g.][]{domiciano_etal14}. On the theoretical
side, much progress has been made recently with the construction of
the first self-consistent (dynamically) two-dimensional (2D) models of
fast rotating stars \cite[e.g.][]{ELR07,ELR13,RELP16}. With these
models it has been possible to make more precise predictions of the
gravity-darkening effect, in particular for rapidly rotating
early-type stars \cite[][]{ELR11,R15}. Currently, interferometric data
and the most recent ESTER models agree very well on the
gravity-darkening exponents \cite[][]{domiciano_etal14}. However, this
is only valid for early-type stars.

For late-type stars the situation is less clear. As pointed out above,
\cite{lucy67} was the first to propose a theoretical estimate of
gravity darkening for late-type stars. He actually suggested that
$\beta\approx 0.08$ for main sequence stars with masses approximately
equal to the solar mass.  However, as shown in \cite{ELR12}, Lucy's
approach leads to a gravity-darkening exponent that is essentially
controlled by the opacity law in the surface layers and does not
reflect the effects of the expected anisotropies of the underlying
rotating convection. Interferometric data from the star $\beta$ Cas,
which is beyond the main sequence and most likely owns a convective
envelope, point to $\beta\approx 0.14$ \cite[][]{che_etal11}, thus
also requiring a new modelling.

However, modelling the latitude dependence of the heat flux in a fast
rotating late-type star is a thorny problem. Basically, three effects
combine and potentially modulate the heat flux \cite[][]{R15}. The
first, which is expected to be the most important one, is the effect
of the Coriolis acceleration. It tends to make the flows in a columnar
shape, with columns parallel to the rotation axis, inhibiting
convection near the pole and favouring it near the equator, thus
pointing to a negative gravity-darkening exponent. The second effect
is the centrifugal effect that diminishes the buoyancy in the
equatorial regions and thus contributes to a positive
gravity-darkening exponent.  Finally, fluid flows generate magnetic
fields that can also inhibit heat transfer, both in the bulk or at the
surface via spots.

The above arguments show that modelling gravity darkening for
stars possessing a convective envelope is far from easy. To make a step
forward in this modelling, we investigate here the latitudinal
variations of the flux at the surface of a fluid contained in a
rotating spherical shell and heated from below. To that end, we
perform direct numerical simulations using the anelastic
approximation (sound waves are filtered out but background density
variations are taken into account). As a first step, we concentrate
solely on the Coriolis effect. Thus centrifugal and dynamo effects are
neglected. They will be implemented and investigated in the subsequent
studies.

The paper is organised as follows: Section~\ref{s:model} introduces the
anelastic models and the numerical solvers used for this study.
Results are presented in Sect.~\ref{s:results} and discussed in
Sect.~\ref{s:discussion}. Finally, a set of critical Rayleigh numbers
for the linear onset of convection and the overview of the numerical
simulations carried out are given in Appendices~\ref{app:rac} and
\ref{app:models}, respectively.

\section{Modelling}\label{s:model}


We consider a spherical shell in rotation at angular
velocity~$\Omega\,\ez$, bounded by two concentric spheres of
radius~\ri{} and \ro{}, and filled with a perfect gas with kinematic
viscosity~$\nu$, turbulent entropy diffusivity~$\kappa,$ and specific
heat~$c_p$ (all taken as constants).  Independently of the shell
aspect ratio~$\aspectratio = r_\text{i}/r_\text{o}$, we assume that
the mass bulk is concentrated inside the inner surface~\ri{} and we
further neglect the centrifugal acceleration, which results in the
radial gravity profile ${\vect{g}}=-GM\er/r^2$, where $G$ is the
gravitational constant and $M$ the central mass. The fluid flow is
modelled using the LBR anelastic equations, named after
\citet{braginsky95} and \citet{lantz99}. Our set-up is actually
equivalent to the one used in the anelastic dynamo benchmark
\citep{jones11}, in which the closure relation for the heat flux is
expressed in terms of the entropy gradient.  Convection is then driven
by an imposed entropy difference $\Delta S$ between the inner and
outer boundaries. In the following, we recall the equations for
completeness and refer the reader to \cite{wood2016} for a discussion
of the definition of consistent thermodynamic variables in sound-proof
approximations of the Navier-Stokes equation \citep[see also][]{calkins2015}.

\begin{figure*}
  \centering
  \includegraphics[width=0.49\textwidth]{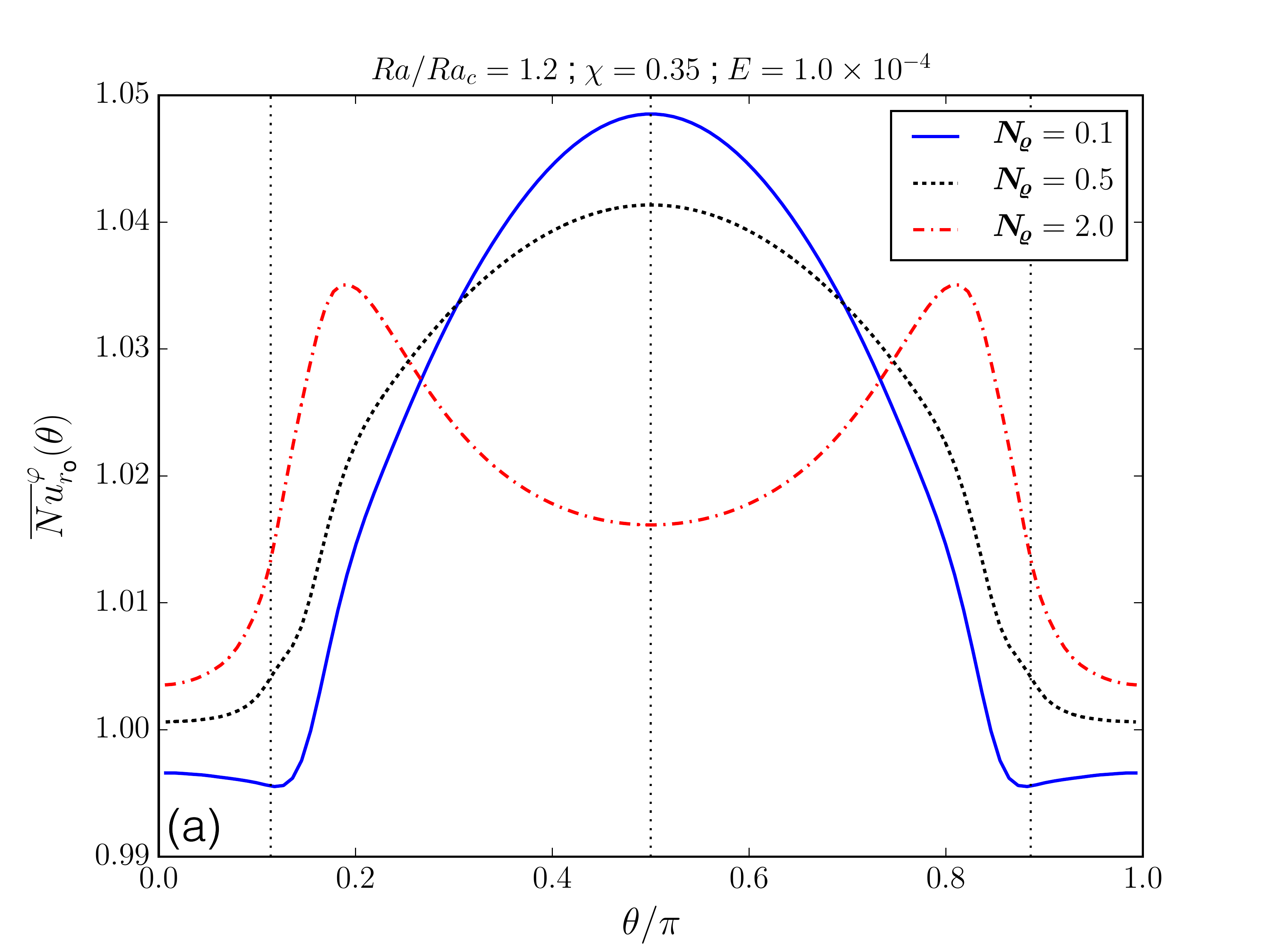}
  \includegraphics[width=0.49\textwidth]{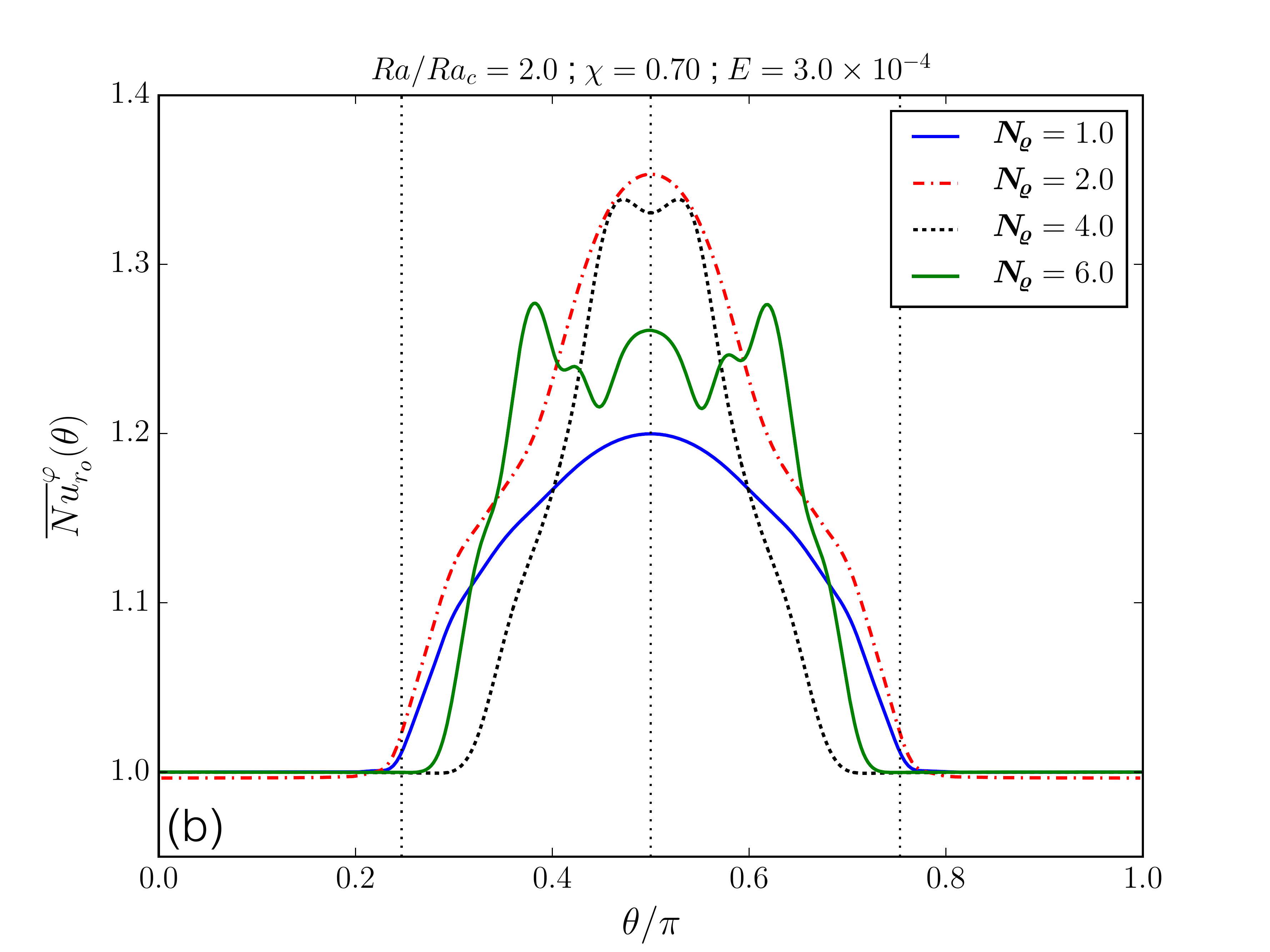}
  \caption{Nusselt number as a function of colatitude close to the
    onset of convection with increasing density stratification, for
    thick (a) and thin (b) shells. The positions of the equator and
    the tangent cylinder are indicated by vertical dotted
    lines.}\label{f:onset}
\end{figure*}

In our models, the reference state is the polytropic solution of the
hydrostatic equations for an adiabatically stratified atmosphere,
which reads
\begin{equation}
  \mT=T_c\, \zeta\,,\qquad \mrho=\varrho_c\,
  \zeta^\npol\,,\qquad \mP=P_c\,
  \zeta^{\npol+1}\,,\label{eq-ref}
\end{equation}
with
\begin{align}
  \zeta&=c_0+c_1 d/r \,,  
  &c_0&=\frac{2\zeta_\text{o}-\aspectratio-1}{1-\aspectratio}\,,\\
  c_1&=\frac{(1+\aspectratio)(1-\zeta_\text{o})}{(1-\aspectratio)^2} \,,
  &\zeta_\text{o}&=\frac{\aspectratio+1}{\aspectratio\exp(\Nrho /\npol)+1} 
  \,.\label{eq:zeta}  
\end{align}
The constants $P_c$, $\varrho_c$ and $T_c$ in Eq.~\eqref{eq-ref} are
the reference-state pressure, density, and temperature midway between
the inner and outer boundaries.  These reference values serve as units
for these variables, whilst length is scaled by the shell
width~$d=r_\text{o}-r_\text{i}$, time by the viscous time~$d^2/\nu,$
and entropy by the entropy drop~$\Delta S$. Then, the coupled
Navier-Stokes and heat transfer equations take the form
\begin{align}
    \frac{\partial \vect{v}}{\partial t} +
    \left(\vect{v}\cdot\bnabla\right) \vect{v} &=
    -\frac{1}{\Ekman}\bnabla\left(\frac{P'}{\zeta^n}\right)
    +\frac{\Ra}{\Prandtl}\frac{s}{r^2} \er -\frac{2}{\Ekman}\,\ez
    \times\vect{v} + \vect{F}_\nu \,, \label{eq:ns}\\ \frac{\partial
      S}{\partial t} + \left(\vect{v}\cdot\bnabla\right) S &=
    \frac{\zeta^{-\npol-1}}{\Prandtl}\bnabla\cdot\left(\zeta^{\npol+1}\,\bnabla
    S\right) + \frac{\Di}{\zeta} Q_\nu \,,\label{eq:ht}\\ \bnabla\cdot
    \left(\zeta^n \vect{v} \right) &= 0 \,.\label{eq:mass}
\end{align}
The control parameters of the above system are:
\begin{align}
  &\text{the Rayleigh number}\qquad &\Ra&=\frac{GMd\Delta S}{\nu\kappa
    c_p} \,,\label{ra} \\ &\text{the Ekman number}\qquad
  &\Ekman&=\frac{\nu}{\Omega d^2} \,, \\ &\text{the Prandtl number}
  &\Prandtl &= \frac{\nu}{\kappa} \,,\\ &\text{the number of density
    scale heights} &\Nrho &= \ln
  \frac{\mrho(r_\text{i})}{\mrho(r_\text{o})}
  \,,
\end{align}
together with the shell aspect ratio~\aspectratio{} and the polytropic
index~\npol{}.  In Eq.~\eqref{eq:ns}, $P'$ denotes the pressure
perturbation and the viscous force~$\vect{F}_\nu$ is given by
$\vect{F}_\nu=\zeta^{-n}\bnabla\tensor{S}$, where the rate of strain
tensor~\tensor{S} is defined by

\begin{equation}
  \tensor{S}_{ij}=2\zeta^n\left(e_{ij}-\frac{1}{3}
  \tensor{\delta}_{ij}\bnabla\cdot
  \vect{v}\right) \quad\text{and}\quad
  e_{ij}=\frac{1}{2}\left(\partial_j v_i+ \partial_i v_j\right) \,,
\end{equation}
since the kinematic viscosity is assumed to be constant.  The viscous
heating~$Q_\nu$ is then given by
\begin{equation}
  Q_\nu=2\left[e_{ij}e_{ji}-\frac{1}{3}(\bnabla\cdot\vect{v})^2\right] 
  \,.
\end{equation}
Finally, the expression of the dissipation parameter~\Di{} in
Eq.~\eqref{eq:ht} reduces to 
\begin{equation}
  \Di= \frac{\nu^2}{d^2 T_c \Delta S}=c_1 \frac{\Prandtl}{\Ra} 
  \,,
\end{equation}
where the last equality follows from the hydrostatic balance $\grad
\mP=\mrho\vect{g}$ and the equation of state of an ideal gas close to
adiabatic, $\mP =\mrho \mT c_p/(n+1)$.

Since we are primarily interested in modelling stellar convection
zones, we impose impenetrable and stress-free boundary conditions for
the velocity field,
\begin{equation}
  v_r = \frac{\partial}{\partial r}\left( \frac{v_\theta}{r}\right) =
  \frac{\partial}{\partial r}\left( \frac{v_\varphi}{r}\right) = 0
  \quad\text{on}\quad r=\ri \quad\text{and}\quad r=\ro
  \,,
\end{equation}
whereas the entropy is fixed at the inner and outer boundaries.
Stress-free conditions are justified even at the bottom of the
convection zone since the (turbulent) viscosity of the convection zone
is much larger than the viscosity of the radiative region
\citep{ruediger1989,rieutord2008}.

The time integration of the anelastic
system~\eqref{eq:ns}--\eqref{eq:mass} has been performed with two
different pseudo-spectral codes, \parody{}
\citep{dormy98,schrinner2014} and \magic{}\footnote{\magic{} is
  available online at \url{http://github.org/magic-sph/magic}. It uses
  the SHTns library available at
  \url{https://bitbucket.org/nschaeff/shtns}.}
\citep{gastine12a,schaeffer2013}.  These codes both use a
poloidal-toroidal decomposition to ensure the solenoidal
constraint~\eqref{eq:mass}, the major difference lying in the radial
discretization: \parody{} is based on a finite difference scheme,
while \magic{} uses Chebyshev polynomials. Both numerical solvers
reproduce the anelastic dynamo benchmark \citep{jones11} and we
checked on a test case that the results we obtain do not differ from
one solver to another.  In practice, one must be careful
that the default definition of the Rayleigh number in \magic{}
slightly differs from the one given in Eq.~\eqref{ra} and obeys the
relation $\RaM = \Ra \left(1-\aspectratio\right)^2$.
 
The reader will find a set of critical Rayleigh number values for the
linear onset of convection in Table~\ref{t:rac}. These have been
calculated solving the eigenvalue problem of the linearized equations
of perturbations as in \cite{jones_etal09}, using a spectral
decomposition on the spherical harmonics and Chebyshev polynomials
together with an Arnoldi-Chebyshev solver
\cite[][]{RV97,VRBF07}. Other critical Rayleigh numbers may be found
in \citet{schrinner2014}. Table~\ref{t:models} contains the summary of
the numerical models and specifies their integration time $\Delta t$
(expressed in turnover time units $d/v^\text{nz}_\text{rms}$ computed
with the non-zonal velocity field) and their spatial
resolution. Numerical convergence has been empirically checked on the
basis of a decrease of at least two orders of magnitude in the kinetic
energy spectra; although this criterion is not always sufficient to
prevent spurious fluctuations of the Nusselt number at the poles.
This latter is defined as the output luminosity normalised by the
conductive state luminosity and its expression at the outer surface
reduces to $\Nu\left(\ro,\theta,\varphi\right) = (\er\cdot \grad S)
/(\er\cdot\grad \Scond)$, with the conductive entropy profile~\Scond{}
\begin{equation}
  \Scond\left(r\right) = \frac{\zeta_{\text{o}}^{-n } -
    \zeta^{-n}}{\zeta_\text{o}^{-n } - \zeta_\text{i}^{-n }} 
  \,,\label{eq:scond}
\end{equation}
where $\zeta_\text{o}$ is given by Eq.~\eqref{eq:zeta} and
$\zeta_\text{i}=(1+\aspectratio - \zeta_\text{o})/\aspectratio$.  The
above equations result in the distribution of the surface heat flux as
a function of colatitude~$\theta$ given by
\begin{equation}
  \overline{Nu}^{\varphi}_{\ro} \left(\theta\right) = - \frac{\left(1
    - e^{-\Nrho} \right) \zeta_{\text{o}} \ro^2 }{n c_1}
  \frac{1}{2\pi}\int_0^{2\pi}\left.\frac{\partial S}{\partial r}
  \right|_{\ro} \dif \varphi \,.
\end{equation}
In the bulk, we must account for the heat advected by the flow, which
gives

\begin{equation}
  \overline{Nu}^{\varphi}(r,\theta) =
  \frac{(1-e^{-\Nrho})\zeta^{\npol+1}r^2}{\npol c_1 \zeta_o^{\npol}}
  \frac{1}{2\pi}\int_0^{2\pi} \left(\Prandtl S' u_r - \frac{\partial
    S}{\partial r} \right) \dif \varphi \,,
\end{equation}
with the entropy perturbation $S'=S-\Scond$.

\section{Results}\label{s:results}

\begin{figure*}
  \centering
  \includegraphics[width=17cm]{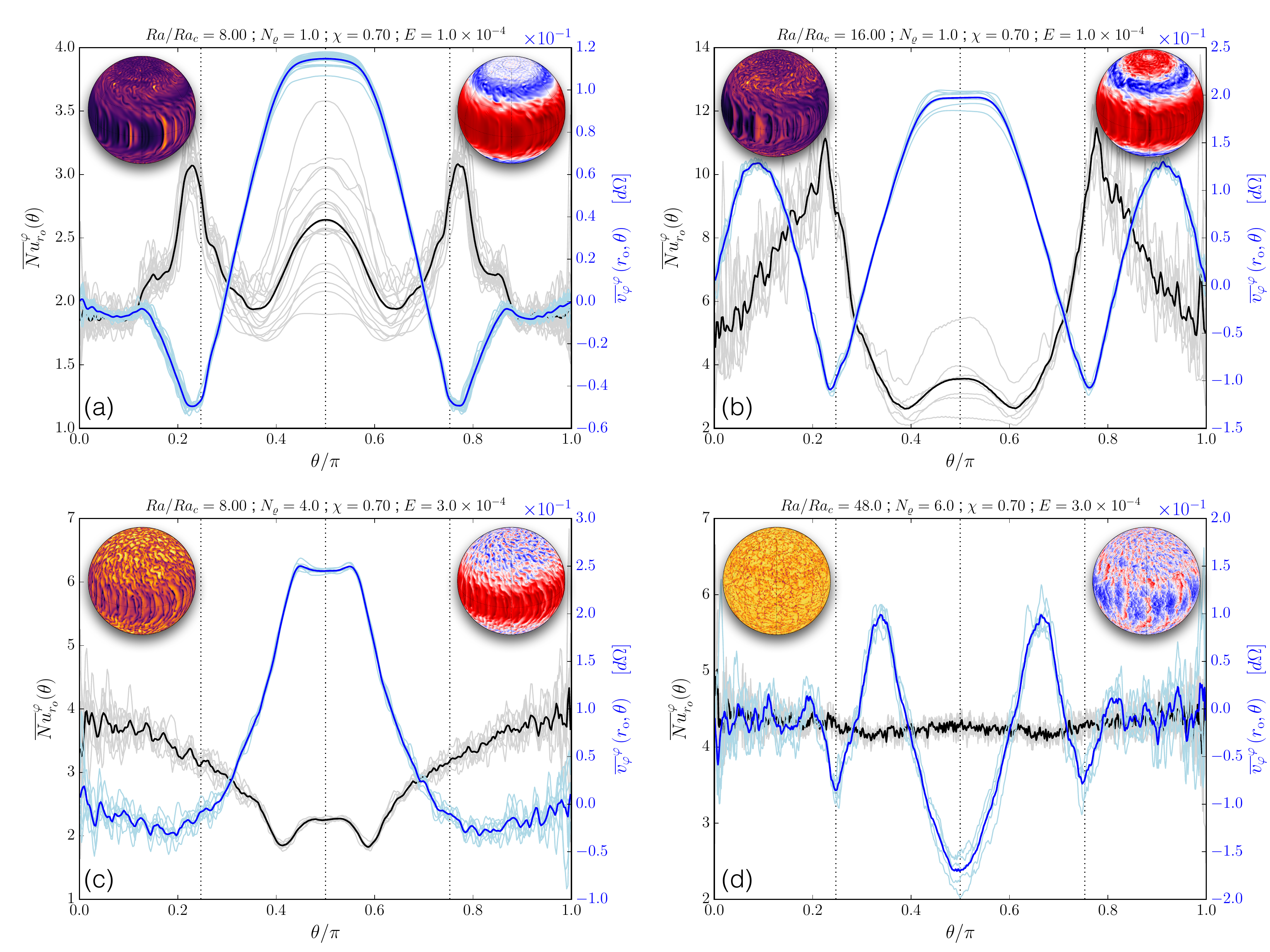}
  \caption{Nusselt (black) and zonal velocity (blue) profiles as a
    function of colatitude for different thin shell models. The colour
    insets represent snapshots of $S(r=0.98\ro)$ and
    $v_\varphi(r=\ro)$. The positions of the equator and the tangent
    cylinder are indicated by vertical dotted lines.}\label{f:prof}
\end{figure*}

For a Boussinesq fluid characterised by a Prandtl number of approximately unity,
it is well known that the onset of convection driven by differential
heating in a rotating spherical shell takes the form of columns
aligned with the rotation axis, sometimes described in terms of
quasi-geostrophic, eastward travelling thermal Rossby waves
\citep{busse1970,jones2000,dormy2004}.  In this regime, the axially
aligned convective rolls do not break the equatorial symmetry and
transfer heat preferentially in the direction perpendicular to the
rotation axis, whereas regions inside the tangent cylinder are nearly
stagnant. The heat flux is then maximum at the equator and symmetric
with respect to the equatorial plane
\citep{tilgner1997,busse06,yadav2016}.  For fluids with a radially
decreasing mean density submitted to differential heating, it has
been shown that compressibility tends to push the convection cells
outward, away from the tangent cylinder
\citep{glatzmaier81_lin,jones_etal09,gastine12a}. We find that this
directly affects the latitudinal heat flux profile and that these
changes are more or less pronounced depending on the shell
thickness. For a thick convective zone ($\aspectratio=0.35$), we see
in Fig.~\ref{f:onset}a that the concavity of the heat flux rapidly
changes with the background density contrast. As expected, the heat
flux is maximum at the equator for a quasi-Boussinesq set-up with
$\Nrho=0.1$ (solid blue line). As \Nrho{} increases, it tends to
flatten ($\Nrho=0.5$, dotted black line) and eventually peaks at two
different latitudes ($\Nrho=2$, red dashed line) which are located
closer to the colatitudes~$\theta_c^\pm$ where the tangent cylinder
crosses the outer surface, determined by $\theta_c^+
=\arcsin\aspectratio$ and $\theta_c^-=\pi-\theta_c^+$.
For a solar-like convective zone ($\aspectratio=0.7$), the surface of
the polar caps that lie inside the tangent cylinder increases,
representing 29~\% of the total outer surface, whereas it is only 6~\%
when $\aspectratio=0.35$. Consequently, the impact of the density
stratification on the heat flux profile at the onset of convection
gets more confined on either side of the equator, as we can see in
Fig.~\ref{f:onset}b.

When we depart further from the onset, convective motions
progressively develop in the tangent cylinder and we observe the
emergence of heat flux maxima located {inside} the tangent
cylinder. This is a general tendency, independent of other parameters
like the shell aspect ratio, the density stratification or the Ekman
number. An example of this transition is given in Fig.~\ref{f:prof}a,
which shows the averaged Nusselt profile (solid black line) for a thin
shell with moderate stratification.  For $\Ra/\Rac=8$, the heat flux
profile displays three distinct maxima as a function of latitude. The
local one that is centred on the equator corresponds to the
equatorial maximum we observe at the onset of convection in thin
shells. Its average value is about
$\overline{Nu}^{\varphi}_{\ro}(\pi/2)\approx 2.5$ and has increased by
25~\% when compared to its value for $\Ra/\Rac=4$. The pair of
absolute maxima located at the boundary of the tangent cylinder do not
exist for $\Ra/\Rac=4$. When we double again the Rayleigh number to
reach $Ra/\Rac=16$, we switch from this intermediate state to a
situation where the surface heat flux is predominantly concentrated
inside the tangent cylinder, as we can see in Fig.~\ref{f:prof}b. The
contrast between the equatorial heat flux and the heat flux at the
tangent cylinder has more than doubled. However, we still distinguish
a local maximum at the equator with almost the same value, surrounded
by two dips at $\theta/\pi \approx 0.4$ and $\theta/\pi \approx 0.6$. When
comparing Figs.~\ref{f:prof}a and \ref{f:prof}b, we note that the
latitudes of the extrema are identical in both cases. A similar
situation prevails in thick shells, except that we do not observe any
extrema in the equatorial belt where the heat flux profile tends to be
flat for $\theta/\pi \in \left[0.4, 0.6\right]$. In this regime, the
signature of the tangent cylinder is then the characteristic feature
of the latitudinal variations of the Nusselt number.

For higher density stratification, we observe a similar regime where
the heat flux is stronger inside the tangent cylinder, with the
difference that the Nusselt number tends to be maximum at the poles
but not anymore at the tangent cylinder boundaries, as illustrated in
Fig.~\ref{f:prof}c. Moreover, increasing both the density
stratification and the Rayleigh number leads us to the discovery of a
third type of Nusselt profile almost constant in latitude. This regime
strongly differs from previous observations and seems typical of
turbulent, large \Nrho{} models. The limit of high \Ra{} and high
\Nrho{} is, of course, very difficult to achieve numerically, but our
set of simulations indicates that the heat flux has a tendency to
flatten for high enough Rayleigh numbers and ultimately become
independent of the latitude, as shown in the example displayed in
Fig.~\ref{f:prof}d. In both thin and thick shells, the flattening of
the heat flux profile results from the decrease of the contrast
between the equatorial and polar heat fluxes. It is relatively smooth
and for this reason, it is sometimes difficult to arbitrarily
distinguish this regime from the previous one. Nevertheless, we notice
that the evolution toward a homogeneous surface heat flux is favoured
by high density contrasts, since we did not observe such flat profiles
for $\Nrho\leq2$ for $\Ra/\Rac\leq 40$.  The larger \Nrho{}, the
faster we reach this regime when increasing the Rayleigh number.

\section{Discussion}\label{s:discussion}

\begin{figure}
  \resizebox{\hsize}{!}
            {\includegraphics{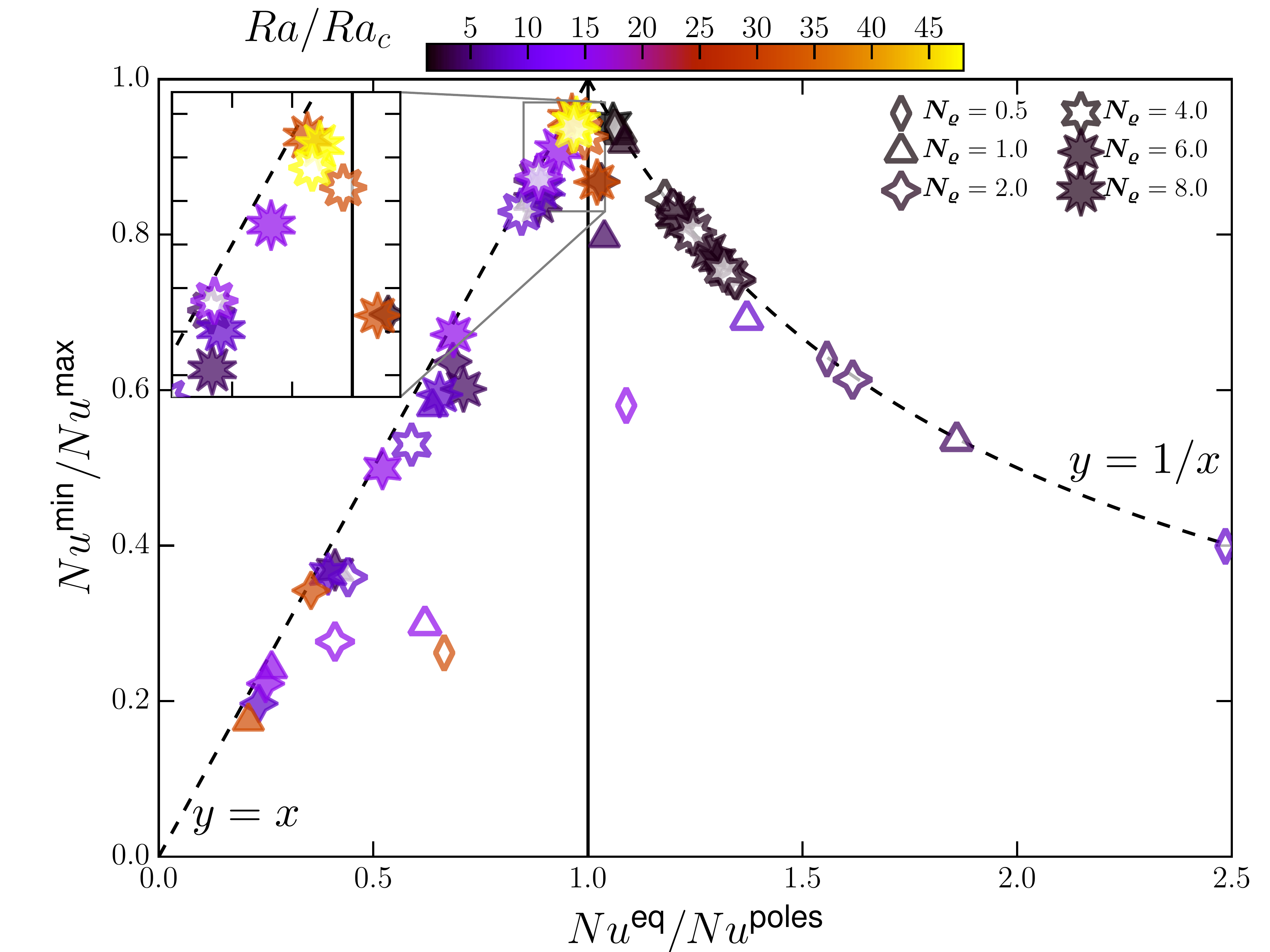}}
  \caption{Ratio $\min \overline{Nu}^{\varphi}_{\ro} (\theta) / \max
    \overline{Nu}^{\varphi}_{\ro}(\theta)$ as a function of the ratio
    $\overline{Nu}^{\varphi}_{\ro}
    (\text{eq})/\overline{Nu}^{\varphi}_{\ro}(\text{poles})$ for our
    sample of models. The symbol shape and colour indicate the number
    of density scale heights and the departure from the onset,
    respectively. Empty/full symbols are used for thin/thick shell
    models.}\label{f:minmax}
\end{figure}

Our systematic parameter study reveals that the variations in latitude
of the heat flux transported by convection at the surface of a
rotating spherical shell strongly vary in the parameter space.
Various profiles have indeed been identified, ranging from brighter to
darker equator or uniform profiles. These different regimes can be
identified in Fig.~\ref{f:minmax}, which displays the ratio $\min
\overline{Nu}^{\varphi}_{\ro} (\theta) / \max
\overline{Nu}^{\varphi}_{\ro}(\theta)$ as a function of the ratio
$\overline{Nu}^{\varphi}_{\ro}
(\text{eq})/\overline{Nu}^{\varphi}_{\ro}(\text{poles})$. The maximum
and minimum values have been computed after performing a running
average in latitude of the Nusselt profile in order to remove
small-scale fluctuations, and the equator and the polar values have
been averaged on an angular sector of $10\degr$. One can see that the
weakly supercritical models mainly stand on the dashed curve $y=1/x$,
since the equator is usually brighter at the onset of convection. For
$\Ra/\Rac \apprge 10$, the points tend to fall on the dashed line
$y=x$, which indicates that we switch from a brighter to a darker
equator. We also note that the contrast tends to be stronger. Finally,
the third regime is indicated by the group of points that tends to
accumulate close to the intersection of the dashed lines, where models
with $\Nrho\geq 6$ are predominant (see the inset). Of course, we
stress that this representation is too simple to render with precision
all the variations of the Nusselt number that have been observed,
especially when the heat flux is maximum at the tangent cylinder (see
Fig.~\ref{f:prof}, top panels). This peculiar configuration has been
preferentially found for low stratification, which explains why a few
points do not fall on the dashed lines.

\begin{figure}
  \resizebox{\hsize}{!}
            {\includegraphics{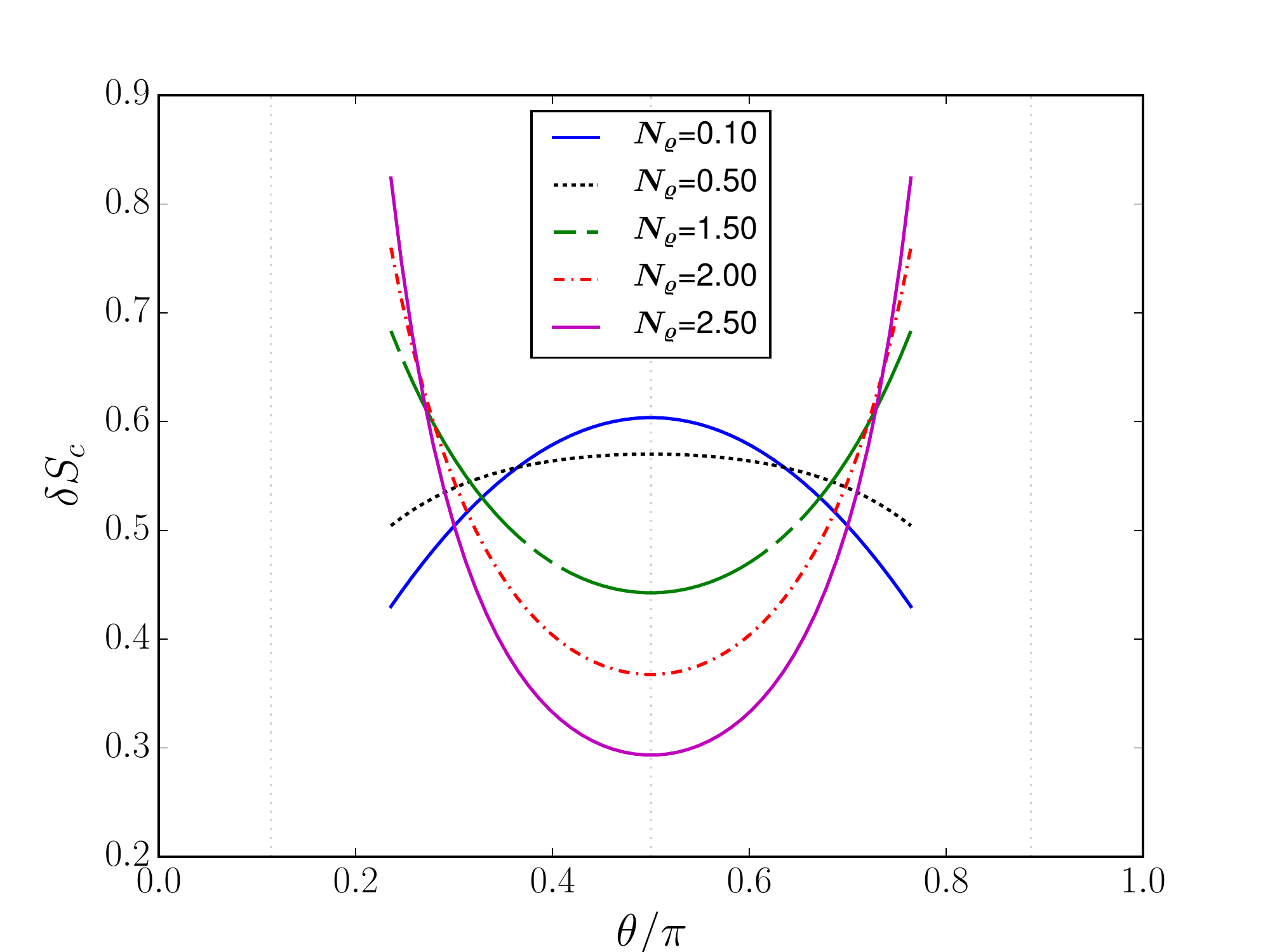}
             \includegraphics[clip=true, trim= 40 70 40 45]{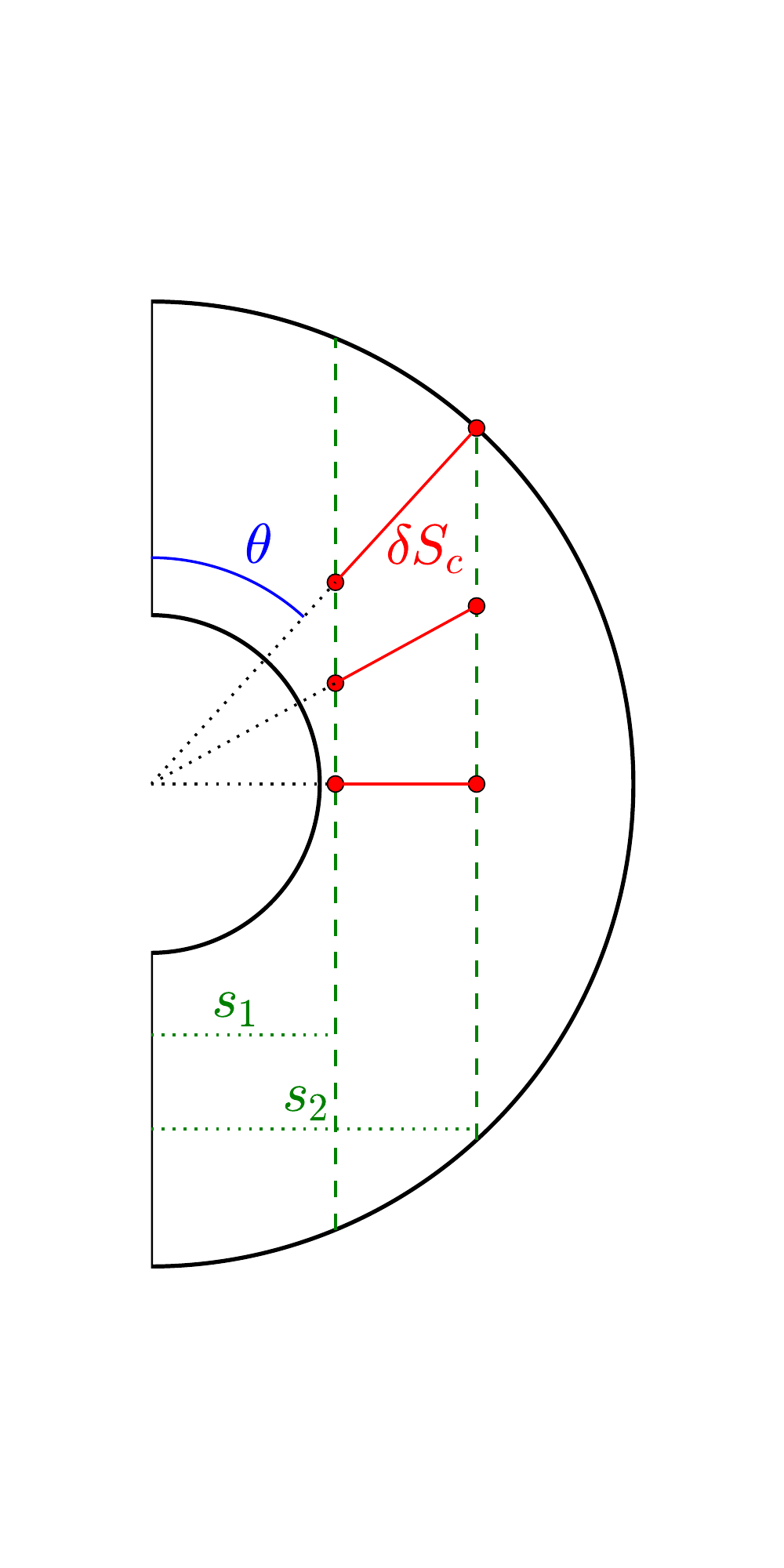}}
  \caption{Left: $\delta \Scond = \Scond(s=s_1) - \Scond(s=s_2)$ as a
    function of colatitude for increasing density
    stratification. Right: sketch illustrating the definition of
    $\delta \Scond$.}\label{f:delta}
\end{figure}

At the onset of convection, one can gain some intuition about the
variations reported in Fig.~\ref{f:onset}a by examining the \Nrho{}
dependence of the radial conductive profile~\Scond{} defined by
Eq.~\eqref{eq:scond}. In order to estimate the latitude at which the
heat transfer will be more efficient at the onset of the convective
instability, one can compute the radial entropy drop~$\delta S_c$
between two cylindrical radii~$s_1$ and $s_2$ that roughly correspond
to the position of the convective columns (sketched out in
Fig.~\ref{f:delta}). It turns out that this quantity displays latitude
variations similar to the Nusselt number profile: when comparing
Figs.~\ref{f:onset}a and \ref{f:delta}, we see that the maximum
location switches from the equator to mid-latitudes as the
stratification increases from $\Nrho=0.1$ to $\Nrho=2$. Thus, the
latitude dependence of the heat flux profile at the onset of
convection is intrinsically linked to the conductive entropy
profile~\Scond{}, which explains why it mainly depends on the shell
aspect ratio and density stratification.

To better understand the regimes shown in Fig.~\ref{f:prof}, it is
interesting to compare the variations in latitude of the Nusselt
number (black lines) to those of the zonal velocity field
$\overline{v_\varphi}^\varphi (\ro, \theta)$ (blue lines). In
Figs.~\ref{f:prof}b and \ref{f:prof}c, the intensity of the zonal wind
is stronger at the equator, which coincides with the position of the
Nusselt minimum.  This observation is consistent with the behaviour
reported for Boussinesq models \citep{aurnou2008,yadav2016} and
clearly illustrates the phenomenon of zonal flows impeding the heat transfer at
low latitudes \citep{goluskin2014}. Then, the observation of a darker
equator results both from the development of a prograde equatorial jet
sustained by Reynolds stresses and from the growth of convective
motions inside the tangent cylinder, where they are less affected by
the zonal flows. Besides, one may notice that the Nusselt maxima in
Fig.~\ref{f:prof}b coincides with the retrograde jets anchored to the
tangent cylinder. These jets, that ensure the conservation of angular
momentum, are typical of stress-free boundary conditions in Boussinesq
simulations \citep{christensen02,aurnou2004}.  We do not observe a
similar differential rotation profile in Fig.~\ref{f:prof}c which only
displays a single equatorial prograde jet and a global heat flux
minimum at the equator.

\begin{figure*}
  \includegraphics[width=0.49\textwidth]{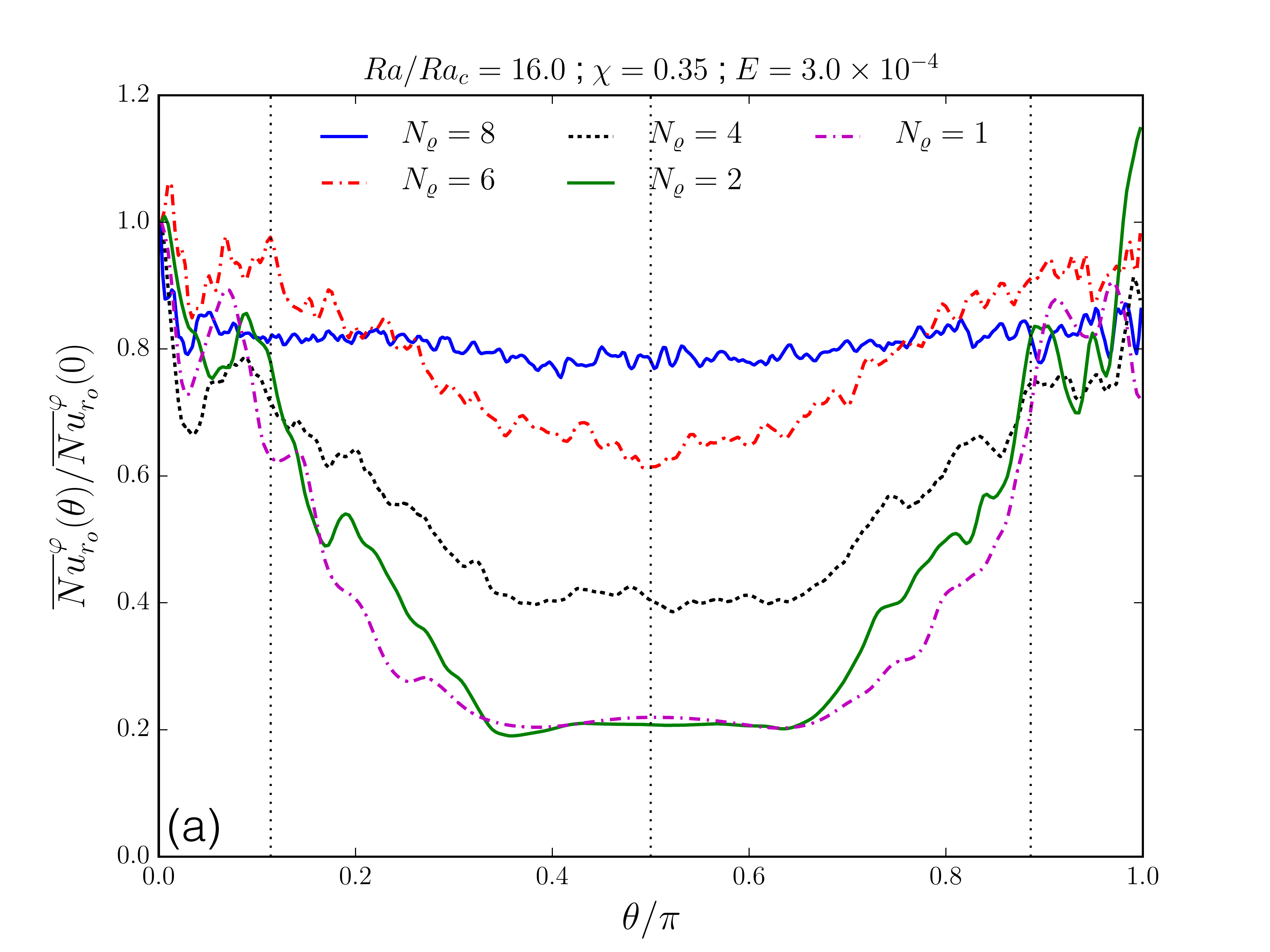}
  \includegraphics[width=0.49\textwidth]{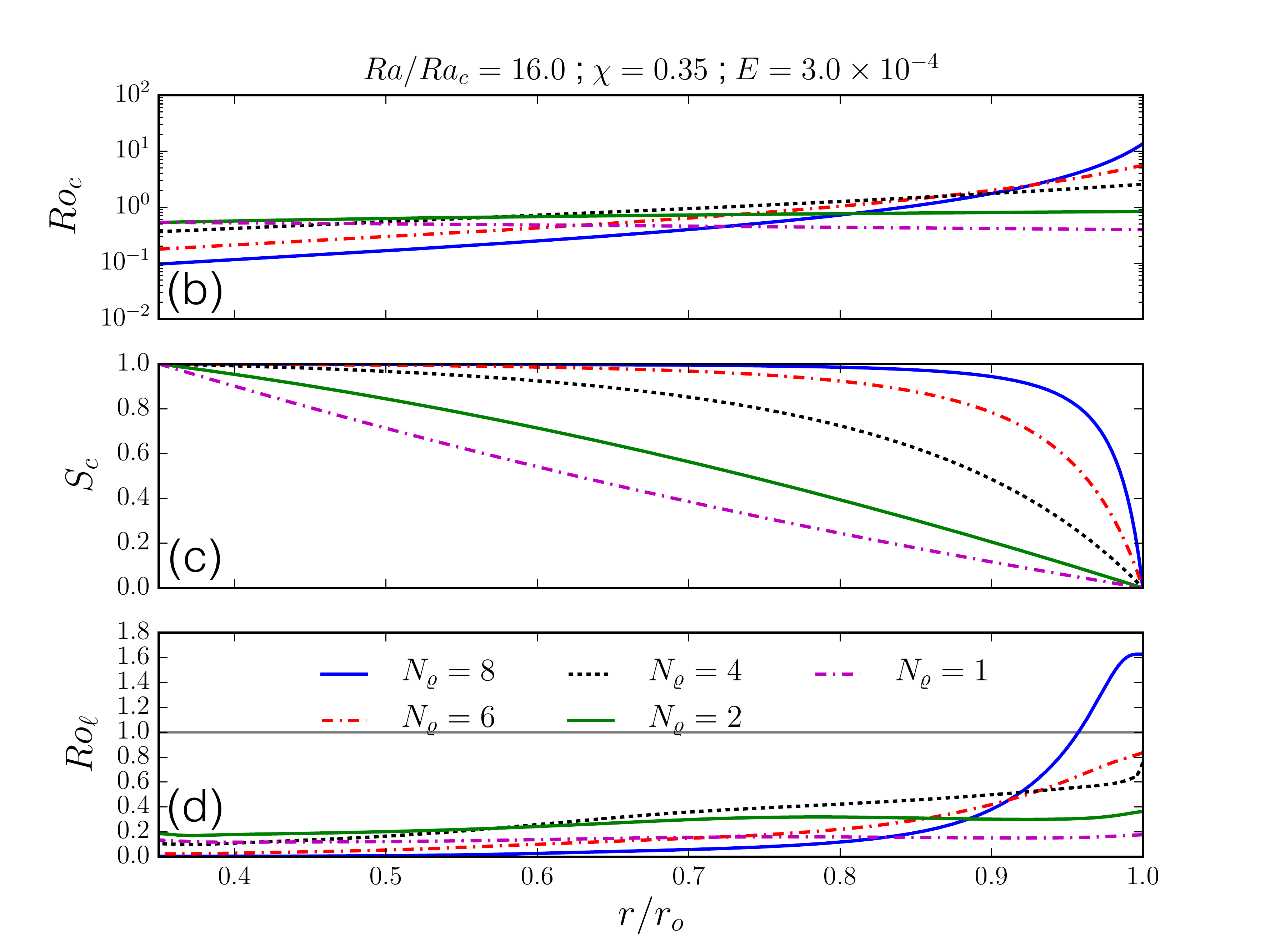}
  \caption{Left: normalised Nusselt profiles (a) averaged in time for
    a subset of thick shell models with decreasing density
    stratification.  Right: radial profiles of the convective Rossby
    number (b), conductive entropy (c) and local Rossby number (d) for
    the same subset of models.}\label{f:sum}
\end{figure*}
\begin{figure*}
  \includegraphics[width=\textwidth]{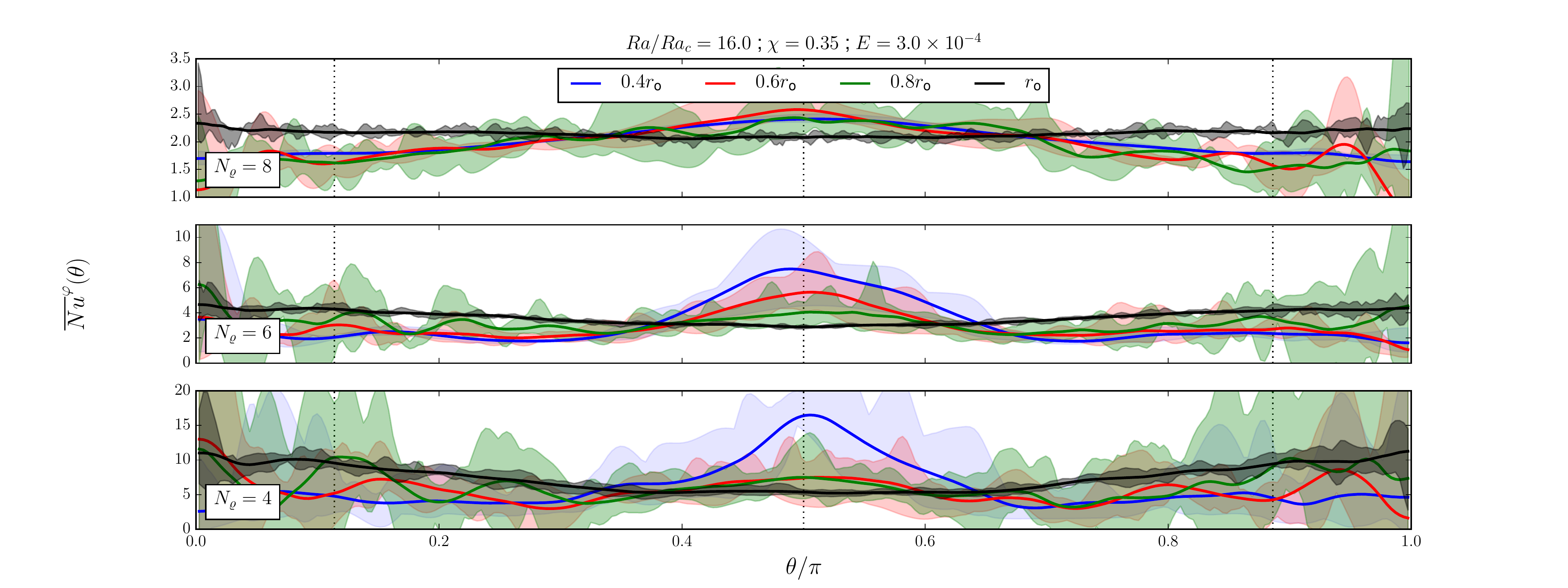}
  \caption{Nusselt profiles as a function of colatitude at different
    depths, for different density stratification. A running average in
    latitude has been applied to the mean profiles (solid lines) and
    the shaded areas highlight the fluctuation
    envelopes.}\label{f:deep}
\end{figure*}

However, it is important to note that this correlation between zonal
wind and heat flux profiles weakens for models with higher density
contrasts, as we can see in the example given in Fig.~\ref{f:prof}d:
the zonal wind is now retrograde at the equator while the heat flux
profile has become almost constant in latitude. As we mention above, this
flattening of the heat flux profile is characteristic of strongly
stratified models, but it appears independent of the specific nature
of the differential rotation profile (solar-like or antisolar).  Since
in the limit $\Omega \to 0$ the system is expected to recover a
central symmetry and a Nusselt number invariant in latitude, we
believe that the homogenisation of the heat flux results from the
relative diminution of the Coriolis force in the outer layers of the
fluid shell.  Indeed, only strongly stratified anelastic models can
exhibit different dynamical regimes that coexist inside the convective
zone, due to the important variation of the force balance as a
function of depth.  In a first approximation, \citet{gastine2013}
showed that the radius \rmix{} at which the transition from
rotation-dominated to buoyancy-dominated regimes occurs can be estimated by
solving the equation $\roc(\rmix)=1$, where the convective Rossby
number is defined by
\begin{equation}
  \roc(r)=\sqrt{\frac{g}{c_p\Omega^2}\abs{\frac{d\Scond}{d r}}}
  =\sqrt{\frac{Ra E^2}{Pr}\abs{\frac{d\Scond}{d r}}} 
  \,.
\end{equation}
In what they call the transitional regime, the region above $\rmix$
tends to exhibit three-dimensional, radially oriented convective
structures.  This hydrodynamic transition naturally impacts the
advective component of the Nusselt number which is always dominant
over the conductive one in the bulk. In practice, we see in
Fig.~\ref{f:sum}a the flattening of the Nusselt profile induced by the
increase of the density stratification, while Fig.~\ref{f:sum}b
displays the corresponding increase of convective Rossby number close
to the outer surface. Since the radial dependence of \roc{} is only
determined by the conductive entropy profile shown in
Fig.~\ref{f:sum}c, this explains why uniform heat fluxes prevail at
high~\Nrho{}. Indeed, when the stratification increases, \Scond{}
tends to keep values close to $\Scond(\ri)$ in the bulk and sharply
drops in the outer layers in order to match the surface boundary
conditions.
\begin{figure}
  \includegraphics[width=0.49\textwidth]{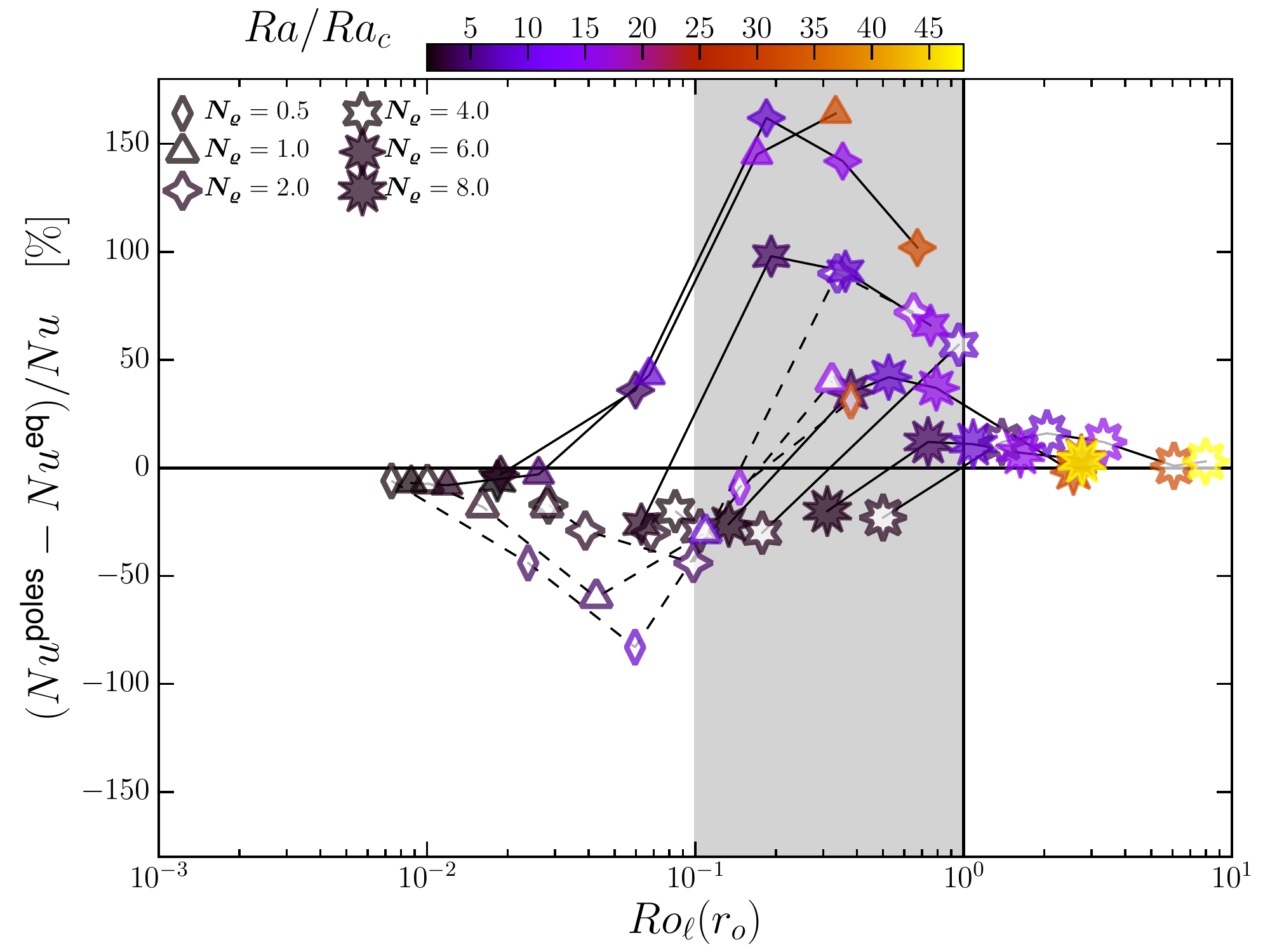}
  \caption{Relative pole/equator contrast as a function of $\rol(\ro)$
    for our sample of models. Solid (dashed) lines indicate
    $E=3\times10^{-4}$ ($E=10^{-4}$) models. The meaning of the
    symbols is defined in the caption of
    Fig.~\ref{f:minmax}.}\label{f:deltanu}
\end{figure}

\begin{figure*}
  \includegraphics[width=0.33\textwidth]{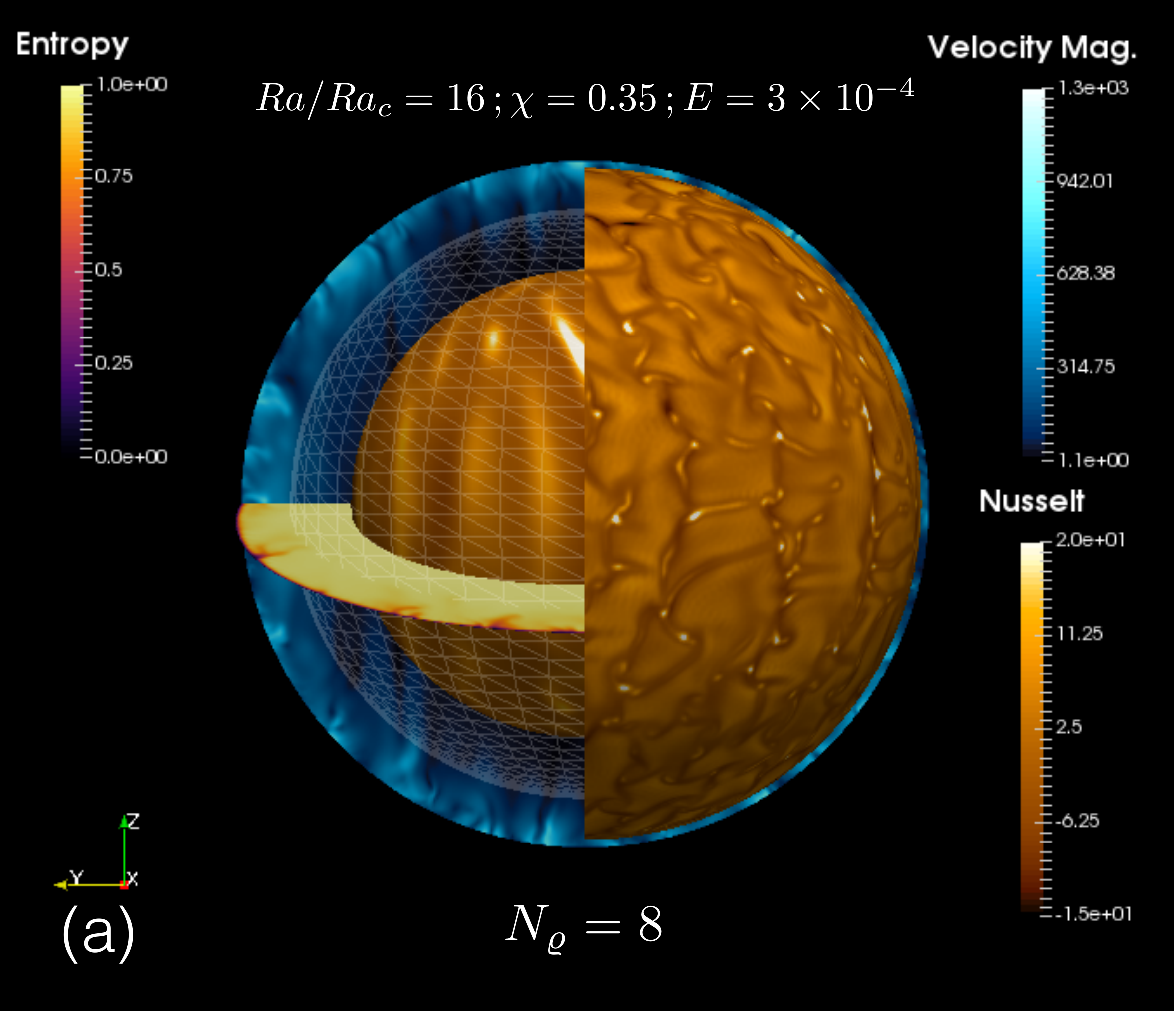}
  \includegraphics[width=0.33\textwidth]{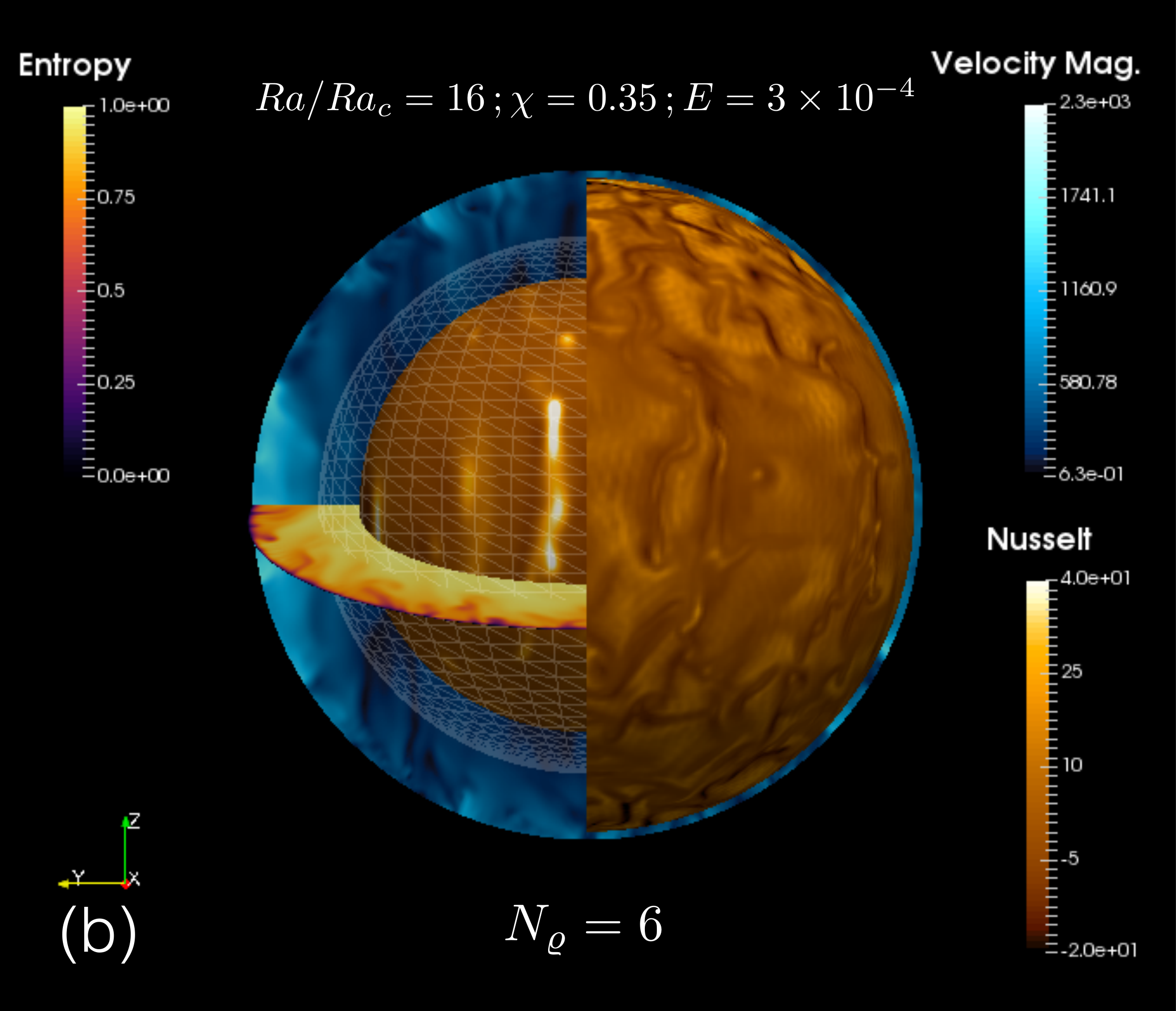}
  \includegraphics[width=0.33\textwidth]{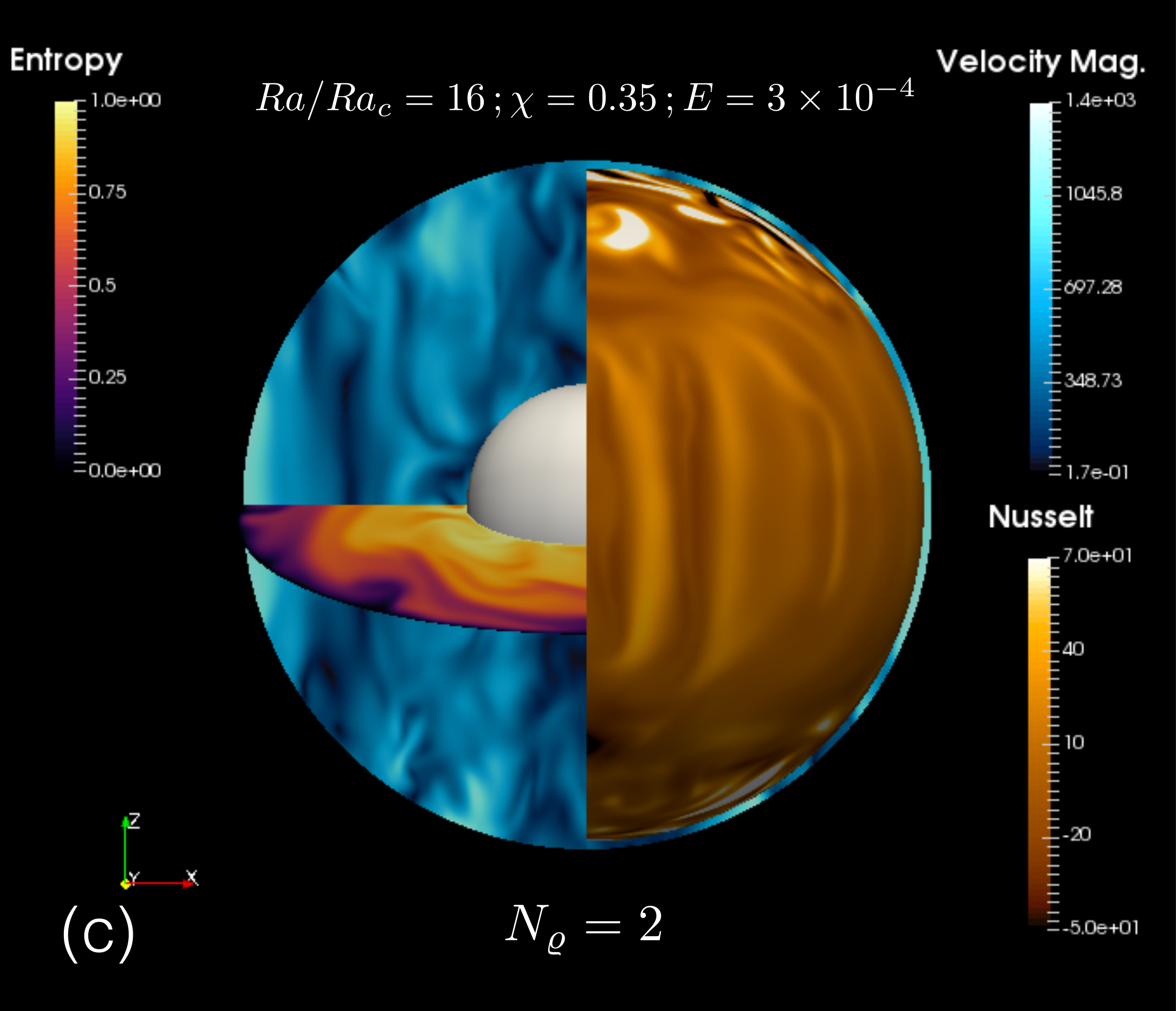}

  \caption{Entropy, velocity magnitude and Nusselt number snapshots
    for decreasing density stratification in thick shell
    models. Nusselt spherical slices have been performed at 0.96\ro{}
    (a--c) and midway between the inner and outer boundaries
    (a,b). Wireframe surfaces (a,b) materialize the radius
    \rmix{}.}\label{f:vtk}
\end{figure*}

However, we stress that a prognostic quantity like \roc{} simply intends
to deliver a rough description of the evolution of the force balance
as a function of depth. A finer estimate of the balance between
inertial and Coriolis forces can be achieved by computing a local
Rossby number $\rol = v_\text{rms}/(\Omega l)$ \citep{christensen06},
where $l=\pi/\ell$ is a typical length scale based on the mean
harmonic degree $\ell$ of the velocity field. We recall that previous
studies highlighted the importance of the mean value of local Rossby
numbers in understanding the field topology of convective dynamos
\citep{christensen06,schrinner12} or the direction of the surface
differential rotation \citep{gastine2014}.  Figure~\ref{f:sum}d
displays the radial profiles of our local Rossby number for different
density stratification. We see that it exceeds unity only in the outer
layers of the model with the highest stratification (blue curve); but
deeper in the bulk, the flow always remains rotationally constrained.
Figure~\ref{f:deep} also shows that the pole/equator luminosity
contrast increases with depth, and that the heat flux tends to be
maximum at the equator deep inside the bulk. As already pointed out by
\citet{durney1981}, this confirms that large pole/equator heat flux
differences in the lower part of the convection zone may coexist with
a negligible luminosity contrast at the surface.

Finally, Fig.~\ref{f:deltanu} shows that the surface value of the
local Rossby number seems to control the heat flux distribution at the
top of the convective zone. If we do not focus on the models close to
the onset of convection, we see that this contrast rapidly decreases
when $\rol(\ro)>1$, whereas it tends to reach its maximum for
$\rol(\ro) \in [0.1, 1]$ when the poles are brighter than the equator
(shaded area). We stress that the collapse of the different models we
see in Fig.~\ref{f:deltanu} would not be obtained using the average
value of the local Rossby number, which tends to be lower than the
surface value for $\Nrho>1$ (the higher the stratification, the lower
the ratio ${\rol}/\rol(\ro)$). This is consistent with the fact that,
in our models, the uniformisation of the heat flux occurs locally,
close to the outer boundary. Although it is theoretically possible to
reach a similar regime with an incompressible model, in practice, the
radial dependence of the conductive entropy profile \Scond{} at large
\Nrho{} is the main cause of the rapid increase of the local Rossby
number in the outer layers of the fluid shell (see
Figs.~\ref{f:sum}c,d).  This impact of the density stratification is
illustrated in Fig.~\ref{f:vtk} which displays snapshot slices of the
entropy, velocity amplitude, and Nusselt number. For the highest
density contrasts (Figs.~\ref{f:vtk}a,b), the velocity amplitude
highlights the transition from rotation-dominated to
buoyancy-dominated flows close to the radius \rmix{}. Moreover, the
outer Nusselt slice in Fig.~\ref{f:vtk}a has been performed just above
the transition $\rol(r)>1$ when the heat flux uniformisation becomes
effective. In contrast, this limit is not achieved in the other panels
with lower stratification.

\section{Conclusion}

As a first numerical approach toward modelling the gravity darkening
in late-type stars endowed with a convective envelope, we carried out
a systematic parameter study to investigate the heat flux distribution
at the surface of a rotating spherical shell filled by a convective
ideal gas.  In the majority of cases, our results are consistent with
the tendencies that have been reported in Boussinesq simulations: at
the onset of convection, the equator is usually brighter, but it
becomes darker than the polar regions when the ratio $\Ra/\Rac$
increases and convective motions fill the tangent cylinder. Favoured
by our choice of stress-free boundary conditions, the equatorial zonal
flow is then efficient at impeding the radial heat transfer at low
latitudes \citep{goluskin2014}.

Besides, thanks to the use of the anelastic approximation, we show
that, among all the system control parameters, the background density
stratification has the strongest impact on the Nusselt number profile.
Indeed, as the stratification increases, the Nusselt number tends to
fluctuate around a constant value in latitude. We show that this
uniformisation of the heat flux distribution turns out to be primarily
controlled by the surface value of the local Rossby number
$\rol(\ro)$, which indicates that it becomes effective in the outer
fluid layers where the Coriolis force is no longer dominating the
dynamics.  In our numerical models, the background density drop and
the shape of the conductive entropy profile \Scond{} at high \Nrho{}
strongly favour the sharp increase of the local Rossby number close to
the outer boundary. This is the reason why we found uniform profiles
only in highly stratified simulations ($\Nrho\geq 6$). In this regime,
the anti-correlation between zonal flows and heat flux which usually
characterises the strongest pole/equator luminosity contrasts
vanishes. Interestingly, we note that the observation of a uniform
energy flux density coexisting with the non-uniform rotation of the
solar surface was at the heart of the so-called heat-flux problem in
theories aimed at explaining the Sun's differential rotation
\citep{ruediger1982}. \citet{rast2008} indeed report a weak
\SI{\approx 0.1}{\percent} enhancement of the solar intensity at
polar latitudes.  The absence of stronger latitudinal variations of
the mean solar photospheric intensity could then be explained by the
fact that convective flows are probably not rotationally constrained
anymore in the near-surface shear layer that spans the outermost 35 Mm
of the Sun \citep{greer2016b,greer2016a}.  \citet{greer2016b} suggest
weak rotational constraint in the outer layers above $r\approx
0.96\ro$, while we find for the thin shell model displayed in
Fig.~\ref{f:prof}d that the transition $\rol=1$ occurs at $r\approx
0.9\ro$ -- a value which is slightly lower than the one predicted from
observations, but we may have deeper transitions in numerical models
given the much lower density stratification of the convective
zone. Moreover, we stress that for this numerical model the radial
profile of the local Rossby number is in very good agreement with the
profile we expect according to the mixing length theory.

In order to connect our results to previous theoretical studies on the
gravity darkening, it would have been interesting to infer
gravity-darkening exponents from our set of direct numerical
simulations. However, by construction, we cannot have access to any
effective gravity $g_\text{eff}$ relying on our numerical models only.
Assuming a Roche model for the surface effective gravity
$g_\text{eff}$, the generalized von Zeipel's law should read $\ln
({Nu^\text{eq}}/{Nu^\text{poles}}) \propto 4 \beta \ln (1+ {\Omega^2
  \ro^3}/({GM}))$, but one has to find a way to estimate the RHS.  For
thin shell models, we attempted to do so using the velocity profile
given by a one-dimensional model of the Sun to derive a rotation rate,
but this approach did not provide a valuable result.

Nevertheless, our study has shown that despite its strength, the
Coriolis force does not seem to be able to break the spherical
symmetry of the exiting heat flux in a rotating star if the local
Rossby number exceeds unity in the surface layers. The short time
scale associated with a short length scale of surface convection
seems to be able to screen the anisotropy of the deep motions of
rotating convection. The natural step forward is now to investigate
the effects of the centrifugal acceleration, which can loosen the
vigour of convection in equatorial regions thanks to reduced
gravity. However, the whole picture might also be strongly perturbed
by magnetic fields. Indeed, \citet{yadav2016} showed with Boussinesq
models that the Nusselt number is enhanced in presence of a magnetic
field which affects the convective motions and tends to quench the
zonal flow at low latitudes. Both effects of magnetic fields and
centrifugal acceleration will be investigated in forthcoming studies.

\begin{acknowledgements}
This study was granted access to the HPC resources of MesoPSL financed
by the R\'{e}gion \^{I}le-de-France and the project Equip@Meso
(reference ANR-10-EQPX-29-01) of the programme Investissements
d'Avenir supervised by the Agence Nationale pour la
Recherche. Numerical simulations were also carried out at the TGCC
Curie and CINES Occigen computing centers (GENCI project A001046698)
as well as at CALMIP -- computing center of Toulouse University (Grant
2016-P1518). R.~R. thanks C.~A.~Jones for comments.
\end{acknowledgements}

\bibliographystyle{aa}
\bibliography{bib/schrinner,bib/raph,bib/bibnew}

\appendix

\section{Critical Rayleigh numbers for the onset of convection}\label{app:rac}

\begin{table}[htbp]
  \caption[]{Critical values: The critical Rayleigh number \Rac{} is
    the one as defined by Eq.~\eqref{ra} (and used in \parody{}); to
    obtain the \magic{} one, just multiply by
    $(1-\aspectratio)^2$. The $L_{\rm max}(10^{-8})$ and
    $N_r(10^{-8})$ quantities give the resolution necessary to achieve
    a relative spectral precision of $10^{-8}$ for the eigenfunction
    at criticality. It gives the minimum resolution needed for the 3D
    simulations.}\label{t:rac} 
  \centering
  \begin{tabular}{cccccc}
    \toprule
    \Nrho & \Rac & $m_c$ & $\omega_c$ & $L_{\rm max}(10^{-8})$ & $N_r(10^{-8})$ \\
    \midrule
    \multicolumn{6}{c}{$\aspectratio=0.35$,  $n=2$, $\Ekman=3\times10^{-4}$}\\
    \midrule
    1 & $ 1.5134 \times 10^5$ & 9  &-174.56 &  60 & 30 \\ 
    2 & $ 3.5094 \times 10^5$ & 12 &-274.86 & 60  & 30 \\ 
    4 & $ 9.9971 \times 10^5$ & 28 &-503.51 & 60  & 30 \\ 
    6 & $ 1.6326 \times 10^6$ & 37 &-863.63 & 65 &  36\\ 
    8 & $ 3.4266 \times 10^6$ & 44 &-1333.1 & 70 & 54 \\
    \midrule
    \multicolumn{6}{c}{$\aspectratio=0.35$,  $\npol=2$, $\Ekman=3\times10^{-5}$}\\
    \midrule
    1.5 & $4.199 \times 10^6$ & 21  &-1167.7 & 90  & 50 \\
    2 & $6.791 \times 10^6$ & 24 &-1376.8 & 120  & 45 \\
    2.5 & $1.053 \times 10^7$ & 29 & - & - & - \\ 
    3 & $1.4771 \times 10^7$ & 44 &-1728.0 & 105  & 50 \\
    \midrule
    \multicolumn{6}{c}{$\aspectratio=0.7$,  $\npol=2$, $\Ekman=3\times10^{-4}$}\\
    \midrule
    1 & $1.659  \times 10^6$ & 36 & -194.34&  90 & 30 \\ 
    2 & $3.134  \times 10^6$ & 54 & -330.83&  92 & 30 \\ 
    4 & $5.573  \times 10^6$ & 74 & -702.57&  92 & 30 \\ 
    6 & $1.1534 \times 10^7$ & 89 &-1173.9& 120 & 50 \\ 
    8 & $3.0645 \times 10^7$ &104 &-1668.7& 130 & 80 \\ 
    \hline
    \noalign{\smallskip}
    \multicolumn{6}{c}{$\aspectratio=0.7$,  $\npol=1.5$, $\Ekman=3\times10^{-4}$}\\
    \midrule
    4   & $6.762 \times 10^6$ & 77 & -801.67& 105 & 30 \\ 
    6   & $1.8828 \times 10^7$ & 92 &-1322.17& 120 & 68 \\ 
    \midrule
    \multicolumn{6}{c}{$\aspectratio=0.7$,  $\npol=2$, $\Ekman=10^{-4}$}\\
    \midrule
    0.5 & $4.154 \times 10^6$ & 43 & -286.78& 120 & 34 \\ 
    1   & $6.790 \times 10^6$ & 52 & -409.7 & 125 & 36 \\ 
    2   & $1.305 \times 10^7$ & 80 & -698.14& 130 & 42 \\ 
    2.5 & $1.507 \times 10^7$ & 90 & -895.47& 134 & 42 \\ 
    4   & $2.107 \times 10^7$ & 110 & -1611.3& 142 & 40 \\ 
    \bottomrule
  \end{tabular}
  \tablefoot{$\Prandtl=1$ for all models. }
\end{table}

\section{Numerical models}\label{app:models}
\onecolumn
\begin{longtable}{ccccccccccccc}
\caption{Numerical simulations carried out at $\Prandtl = 1$ and
  $\npol=2$  and displayed in Figs.~\ref{f:minmax} and
  \ref{f:deltanu}.}\\
\toprule
$ E $&$ \chi $&$ N_{\!\varrho} $&$ Ra/Ra_c $&$ Ro_{\ell}(r_o) $&$ Nu $&$ Nu^\text{min}_{\ro} $&$ Nu^\text{max}_{\ro} $&$ Nu^\text{eq}_{\ro} $&$ Nu^\text{poles}_{\ro} $&$ \Delta t \;[d/v^\text{nz}_\text{rms}] $&$ N_r^\text{max} $&$ \ell_\text{max} $\\
\midrule
\endfirsthead
\caption{(continued)}\\
\hline\hline
$ E $&$ \chi $&$ N_{\!\varrho} $&$ Ra/Ra_c $&$ Ro_{\ell}(r_o) $&$ Nu $&$ Nu^\text{min}_{\ro} $&$ Nu^\text{max}_{\ro} $&$ Nu^\text{eq}_{\ro} $&$ Nu^\text{poles}_{\ro} $&$ \Delta t \;[d/v^\text{nz}_\text{rms}] $&$ N_r^\text{max} $&$ \ell_\text{max} $\\
\hline
\endhead
\bottomrule
\endfoot
$ 1.0 \times 10^{-4} $ & $ 0.35 $ & $ 4.0 $ & $ 1.2 $ & $ 1.8 \times 10^{-2} $ & $ 1.02 $ & $ 1.00 $ & $ 1.06 $ & $ 1.06 $ & $ 1.00 $ & $ 2.3 \times 10^{1} $ & $65$ & $192$ \\
$ 1.0 \times 10^{-4} $ & $ 0.70 $ & $ 0.5 $ & $ 1.5 $ & $ 7.4 \times 10^{-3} $ & $ 1.03 $ & $ 1.00 $ & $ 1.06 $ & $ 1.06 $ & $ 1.00 $ & $ 1.2 \times 10^{0} $ & $129$ & $256$ \\
$ 1.0 \times 10^{-4} $ & $ 0.70 $ & $ 0.5 $ & $ 4.0 $ & $ 2.4 \times 10^{-2} $ & $ 1.25 $ & $ 1.00 $ & $ 1.56 $ & $ 1.56 $ & $ 1.00 $ & $ 3.6 \times 10^{1} $ & $65$ & $213$ \\
$ 1.0 \times 10^{-4} $ & $ 0.70 $ & $ 0.5 $ & $ 8.0 $ & $ 6.0 \times 10^{-2} $ & $ 1.80 $ & $ 1.00 $ & $ 2.51 $ & $ 2.49 $ & $ 1.00 $ & $ 4.7 \times 10^{1} $ & $65$ & $213$ \\
$ 1.0 \times 10^{-4} $ & $ 0.70 $ & $ 0.5 $ & $ 16.0 $ & $ 1.5 \times 10^{-1} $ & $ 3.31 $ & $ 2.61 $ & $ 4.50 $ & $ 3.45 $ & $ 3.17 $ & $ 1.2 \times 10^{2} $ & $129$ & $288$ \\
$ 1.0 \times 10^{-4} $ & $ 0.70 $ & $ 0.5 $ & $ 32.0 $ & $ 3.8 \times 10^{-1} $ & $ 7.51 $ & $ 3.82 $ & $ 14.59 $ & $ 4.63 $ & $ 6.95 $ & $ 1.5 \times 10^{2} $ & $129$ & $341$ \\
$ 1.0 \times 10^{-4} $ & $ 0.70 $ & $ 1.0 $ & $ 1.5 $ & $ 1.0 \times 10^{-2} $ & $ 1.04 $ & $ 1.00 $ & $ 1.08 $ & $ 1.08 $ & $ 1.00 $ & $ 7.7 \times 10^{0} $ & $65$ & $192$ \\
$ 1.0 \times 10^{-4} $ & $ 0.70 $ & $ 1.0 $ & $ 2.0 $ & $ 1.6 \times 10^{-2} $ & $ 1.10 $ & $ 1.00 $ & $ 1.20 $ & $ 1.20 $ & $ 1.00 $ & $ 1.1 \times 10^{1} $ & $65$ & $192$ \\
$ 1.0 \times 10^{-4} $ & $ 0.70 $ & $ 1.0 $ & $ 4.0 $ & $ 4.3 \times 10^{-2} $ & $ 1.43 $ & $ 1.00 $ & $ 1.87 $ & $ 1.86 $ & $ 1.00 $ & $ 1.0 \times 10^{2} $ & $161$ & $256$ \\
$ 1.0 \times 10^{-4} $ & $ 0.70 $ & $ 1.0 $ & $ 8.0 $ & $ 1.1 \times 10^{-1} $ & $ 2.34 $ & $ 1.87 $ & $ 2.71 $ & $ 2.58 $ & $ 1.88 $ & $ 1.4 \times 10^{2} $ & $65$ & $256$ \\
$ 1.0 \times 10^{-4} $ & $ 0.70 $ & $ 1.0 $ & $ 16.0 $ & $ 3.2 \times 10^{-1} $ & $ 5.27 $ & $ 2.89 $ & $ 9.69 $ & $ 3.47 $ & $ 5.60 $ & $ 3.7 \times 10^{2} $ & $129$ & $256$ \\
$ 1.0 \times 10^{-4} $ & $ 0.70 $ & $ 2.0 $ & $ 1.5 $ & $ 2.8 \times 10^{-2} $ & $ 1.07 $ & $ 1.00 $ & $ 1.18 $ & $ 1.18 $ & $ 1.00 $ & $ 1.3 \times 10^{1} $ & $65$ & $192$ \\
$ 1.0 \times 10^{-4} $ & $ 0.70 $ & $ 2.0 $ & $ 2.0 $ & $ 3.9 \times 10^{-2} $ & $ 1.17 $ & $ 1.00 $ & $ 1.35 $ & $ 1.34 $ & $ 1.00 $ & $ 2.7 \times 10^{1} $ & $65$ & $192$ \\
$ 1.0 \times 10^{-4} $ & $ 0.70 $ & $ 2.0 $ & $ 4.0 $ & $ 9.8 \times 10^{-2} $ & $ 1.69 $ & $ 1.19 $ & $ 1.95 $ & $ 1.94 $ & $ 1.20 $ & $ 9.1 \times 10^{1} $ & $65$ & $192$ \\
$ 1.0 \times 10^{-4} $ & $ 0.70 $ & $ 2.0 $ & $ 8.0 $ & $ 3.4 \times 10^{-1} $ & $ 3.78 $ & $ 2.32 $ & $ 6.45 $ & $ 2.67 $ & $ 6.06 $ & $ 5.4 \times 10^{1} $ & $129$ & $256$ \\
$ 1.0 \times 10^{-4} $ & $ 0.70 $ & $ 2.0 $ & $ 16.0 $ & $ 6.5 \times 10^{-1} $ & $ 7.54 $ & $ 3.64 $ & $ 13.16 $ & $ 3.80 $ & $ 9.25 $ & $ 5.7 \times 10^{1} $ & $129$ & $341$ \\
$ 1.0 \times 10^{-4} $ & $ 0.70 $ & $ 4.0 $ & $ 1.5 $ & $ 8.4 \times 10^{-2} $ & $ 1.04 $ & $ 1.00 $ & $ 1.21 $ & $ 1.20 $ & $ 1.00 $ & $ 3.0 \times 10^{1} $ & $65$ & $341$ \\
$ 1.0 \times 10^{-4} $ & $ 0.70 $ & $ 4.0 $ & $ 2.0 $ & $ 1.0 \times 10^{-1} $ & $ 1.08 $ & $ 1.00 $ & $ 1.33 $ & $ 1.32 $ & $ 1.00 $ & $ 4.4 \times 10^{1} $ & $121$ & $341$ \\
$ 3.0 \times 10^{-4} $ & $ 0.35 $ & $ 1.0 $ & $ 1.5 $ & $ 8.7 \times 10^{-3} $ & $ 1.07 $ & $ 1.01 $ & $ 1.08 $ & $ 1.08 $ & $ 1.01 $ & $ 1.8 \times 10^{1} $ & $65$ & $192$ \\
$ 3.0 \times 10^{-4} $ & $ 0.35 $ & $ 1.0 $ & $ 2.0 $ & $ 1.2 \times 10^{-2} $ & $ 1.10 $ & $ 1.02 $ & $ 1.11 $ & $ 1.11 $ & $ 1.02 $ & $ 3.3 \times 10^{1} $ & $65$ & $192$ \\
$ 3.0 \times 10^{-4} $ & $ 0.35 $ & $ 1.0 $ & $ 4.0 $ & $ 2.6 \times 10^{-2} $ & $ 1.36 $ & $ 1.22 $ & $ 1.53 $ & $ 1.28 $ & $ 1.23 $ & $ 2.3 \times 10^{1} $ & $65$ & $192$ \\
$ 3.0 \times 10^{-4} $ & $ 0.35 $ & $ 1.0 $ & $ 8.0 $ & $ 6.7 \times 10^{-2} $ & $ 2.23 $ & $ 1.88 $ & $ 3.25 $ & $ 1.88 $ & $ 2.95 $ & $ 2.8 \times 10^{1} $ & $65$ & $192$ \\
$ 3.0 \times 10^{-4} $ & $ 0.35 $ & $ 1.0 $ & $ 16.0 $ & $ 1.7 \times 10^{-1} $ & $ 4.64 $ & $ 3.14 $ & $ 12.98 $ & $ 3.32 $ & $ 12.62 $ & $ 9.0 \times 10^{1} $ & $65$ & $192$ \\
$ 3.0 \times 10^{-4} $ & $ 0.35 $ & $ 1.0 $ & $ 32.0 $ & $ 3.3 \times 10^{-1} $ & $ 8.90 $ & $ 4.80 $ & $ 27.52 $ & $ 5.37 $ & $ 25.69 $ & $ 4.3 \times 10^{1} $ & $65$ & $192$ \\
$ 3.0 \times 10^{-4} $ & $ 0.35 $ & $ 2.0 $ & $ 2.0 $ & $ 1.9 \times 10^{-2} $ & $ 1.13 $ & $ 1.05 $ & $ 1.21 $ & $ 1.09 $ & $ 1.06 $ & $ 2.1 \times 10^{1} $ & $65$ & $192$ \\
$ 3.0 \times 10^{-4} $ & $ 0.35 $ & $ 2.0 $ & $ 4.0 $ & $ 6.0 \times 10^{-2} $ & $ 1.69 $ & $ 1.34 $ & $ 2.11 $ & $ 1.34 $ & $ 1.96 $ & $ 4.0 \times 10^{1} $ & $65$ & $192$ \\
$ 3.0 \times 10^{-4} $ & $ 0.35 $ & $ 2.0 $ & $ 8.0 $ & $ 1.8 \times 10^{-1} $ & $ 3.64 $ & $ 2.42 $ & $ 12.27 $ & $ 2.48 $ & $ 10.61 $ & $ 4.4 \times 10^{1} $ & $65$ & $192$ \\
$ 3.0 \times 10^{-4} $ & $ 0.35 $ & $ 2.0 $ & $ 16.0 $ & $ 3.5 \times 10^{-1} $ & $ 7.22 $ & $ 4.52 $ & $ 20.32 $ & $ 4.67 $ & $ 18.74 $ & $ 1.3 \times 10^{2} $ & $65$ & $192$ \\
$ 3.0 \times 10^{-4} $ & $ 0.35 $ & $ 2.0 $ & $ 32.0 $ & $ 6.7 \times 10^{-1} $ & $ 13.00 $ & $ 9.00 $ & $ 26.30 $ & $ 9.06 $ & $ 25.52 $ & $ 6.6 \times 10^{1} $ & $129$ & $192$ \\
$ 3.0 \times 10^{-4} $ & $ 0.35 $ & $ 4.0 $ & $ 2.0 $ & $ 6.3 \times 10^{-2} $ & $ 1.21 $ & $ 1.01 $ & $ 1.31 $ & $ 1.31 $ & $ 1.01 $ & $ 5.0 \times 10^{1} $ & $65$ & $192$ \\
$ 3.0 \times 10^{-4} $ & $ 0.35 $ & $ 4.0 $ & $ 4.0 $ & $ 1.9 \times 10^{-1} $ & $ 2.13 $ & $ 1.63 $ & $ 4.41 $ & $ 1.77 $ & $ 4.31 $ & $ 5.8 \times 10^{1} $ & $65$ & $192$ \\
$ 3.0 \times 10^{-4} $ & $ 0.35 $ & $ 4.0 $ & $ 8.0 $ & $ 3.6 \times 10^{-1} $ & $ 3.76 $ & $ 2.64 $ & $ 7.27 $ & $ 2.69 $ & $ 6.81 $ & $ 6.1 \times 10^{1} $ & $65$ & $192$ \\
$ 3.0 \times 10^{-4} $ & $ 0.35 $ & $ 4.0 $ & $ 16.0 $ & $ 7.5 \times 10^{-1} $ & $ 6.39 $ & $ 5.26 $ & $ 10.55 $ & $ 5.36 $ & $ 10.27 $ & $ 8.5 \times 10^{1} $ & $65$ & $192$ \\
$ 3.0 \times 10^{-4} $ & $ 0.35 $ & $ 6.0 $ & $ 2.0 $ & $ 1.3 \times 10^{-1} $ & $ 1.12 $ & $ 1.00 $ & $ 1.28 $ & $ 1.28 $ & $ 1.00 $ & $ 8.8 \times 10^{1} $ & $65$ & $192$ \\
$ 3.0 \times 10^{-4} $ & $ 0.35 $ & $ 6.0 $ & $ 4.0 $ & $ 3.8 \times 10^{-1} $ & $ 1.68 $ & $ 1.38 $ & $ 2.29 $ & $ 1.58 $ & $ 2.22 $ & $ 9.3 \times 10^{1} $ & $65$ & $192$ \\
$ 3.0 \times 10^{-4} $ & $ 0.35 $ & $ 6.0 $ & $ 8.0 $ & $ 5.3 \times 10^{-1} $ & $ 2.42 $ & $ 1.96 $ & $ 3.30 $ & $ 2.12 $ & $ 3.24 $ & $ 6.7 \times 10^{1} $ & $65$ & $192$ \\
$ 3.0 \times 10^{-4} $ & $ 0.35 $ & $ 6.0 $ & $ 16.0 $ & $ 7.9 \times 10^{-1} $ & $ 3.43 $ & $ 2.95 $ & $ 4.39 $ & $ 2.96 $ & $ 4.30 $ & $ 1.4 \times 10^{2} $ & $129$ & $192$ \\
$ 3.0 \times 10^{-4} $ & $ 0.35 $ & $ 6.0 $ & $ 32.0 $ & $ 2.6 \times 10^{0} $ & $ 5.51 $ & $ 5.45 $ & $ 6.29 $ & $ 6.20 $ & $ 6.07 $ & $ 3.6 \times 10^{1} $ & $129$ & $192$ \\
$ 3.0 \times 10^{-4} $ & $ 0.35 $ & $ 8.0 $ & $ 2.0 $ & $ 3.1 \times 10^{-1} $ & $ 1.13 $ & $ 1.00 $ & $ 1.22 $ & $ 1.21 $ & $ 1.00 $ & $ 3.1 \times 10^{1} $ & $129$ & $192$ \\
$ 3.0 \times 10^{-4} $ & $ 0.35 $ & $ 8.0 $ & $ 4.0 $ & $ 7.4 \times 10^{-1} $ & $ 1.44 $ & $ 1.35 $ & $ 1.60 $ & $ 1.38 $ & $ 1.57 $ & $ 2.4 \times 10^{1} $ & $257$ & $288$ \\
$ 3.0 \times 10^{-4} $ & $ 0.35 $ & $ 8.0 $ & $ 8.0 $ & $ 1.1 \times 10^{0} $ & $ 1.73 $ & $ 1.63 $ & $ 1.89 $ & $ 1.63 $ & $ 1.83 $ & $ 2.9 \times 10^{1} $ & $257$ & $288$ \\
$ 3.0 \times 10^{-4} $ & $ 0.35 $ & $ 8.0 $ & $ 16.0 $ & $ 1.6 \times 10^{0} $ & $ 2.12 $ & $ 2.06 $ & $ 2.27 $ & $ 2.09 $ & $ 2.24 $ & $ 3.3 \times 10^{1} $ & $257$ & $288$ \\
$ 3.0 \times 10^{-4} $ & $ 0.35 $ & $ 8.0 $ & $ 32.0 $ & $ 2.7 \times 10^{0} $ & $ 2.76 $ & $ 2.73 $ & $ 2.88 $ & $ 2.75 $ & $ 2.85 $ & $ 3.7 \times 10^{1} $ & $257$ & $288$ \\
$ 3.0 \times 10^{-4} $ & $ 0.35 $ & $ 8.0 $ & $ 48.0 $ & $ 2.8 \times 10^{0} $ & $ 2.91 $ & $ 2.65 $ & $ 2.80 $ & $ 2.65 $ & $ 2.72 $ & $ 4.0 \times 10^{0} $ & $257$ & $512$ \\
$ 3.0 \times 10^{-4} $ & $ 0.70 $ & $ 1.0 $ & $ 2.0 $ & $ 2.8 \times 10^{-2} $ & $ 1.10 $ & $ 1.00 $ & $ 1.20 $ & $ 1.20 $ & $ 1.00 $ & $ 3.4 \times 10^{1} $ & $65$ & $192$ \\
$ 3.0 \times 10^{-4} $ & $ 0.70 $ & $ 2.0 $ & $ 2.0 $ & $ 6.8 \times 10^{-2} $ & $ 1.16 $ & $ 1.00 $ & $ 1.34 $ & $ 1.34 $ & $ 1.00 $ & $ 3.8 \times 10^{1} $ & $65$ & $192$ \\
$ 3.0 \times 10^{-4} $ & $ 0.70 $ & $ 4.0 $ & $ 2.0 $ & $ 1.8 \times 10^{-1} $ & $ 1.10 $ & $ 1.00 $ & $ 1.33 $ & $ 1.33 $ & $ 1.00 $ & $ 2.3 \times 10^{1} $ & $65$ & $192$ \\
$ 3.0 \times 10^{-4} $ & $ 0.70 $ & $ 4.0 $ & $ 8.0 $ & $ 9.6 \times 10^{-1} $ & $ 2.74 $ & $ 2.05 $ & $ 3.88 $ & $ 2.23 $ & $ 3.78 $ & $ 2.5 \times 10^{1} $ & $97$ & $256$ \\
$ 3.0 \times 10^{-4} $ & $ 0.70 $ & $ 6.0 $ & $ 2.0 $ & $ 5.0 \times 10^{-1} $ & $ 1.13 $ & $ 1.00 $ & $ 1.25 $ & $ 1.24 $ & $ 1.00 $ & $ 2.1 \times 10^{1} $ & $65$ & $288$ \\
$ 3.0 \times 10^{-4} $ & $ 0.70 $ & $ 6.0 $ & $ 4.0 $ & $ 1.4 \times 10^{0} $ & $ 1.53 $ & $ 1.44 $ & $ 1.65 $ & $ 1.44 $ & $ 1.64 $ & $ 1.0 \times 10^{1} $ & $65$ & $288$ \\
$ 3.0 \times 10^{-4} $ & $ 0.70 $ & $ 6.0 $ & $ 8.0 $ & $ 2.0 \times 10^{0} $ & $ 1.91 $ & $ 1.77 $ & $ 2.13 $ & $ 1.77 $ & $ 2.10 $ & $ 4.5 \times 10^{0} $ & $65$ & $426$ \\
$ 3.0 \times 10^{-4} $ & $ 0.70 $ & $ 6.0 $ & $ 16.0 $ & $ 3.3 \times 10^{0} $ & $ 2.56 $ & $ 2.46 $ & $ 2.82 $ & $ 2.47 $ & $ 2.80 $ & $ 2.1 \times 10^{1} $ & $129$ & $426$ \\
$ 3.0 \times 10^{-4} $ & $ 0.70 $ & $ 6.0 $ & $ 32.0 $ & $ 6.1 \times 10^{0} $ & $ 3.45 $ & $ 3.37 $ & $ 3.64 $ & $ 3.56 $ & $ 3.58 $ & $ 4.4 \times 10^{0} $ & $129$ & $512$ \\
$ 3.0 \times 10^{-4} $ & $ 0.70 $ & $ 6.0 $ & $ 48.0 $ & $ 8.0 \times 10^{0} $ & $ 4.31 $ & $ 4.18 $ & $ 4.47 $ & $ 4.26 $ & $ 4.41 $ & $ 3.7 \times 10^{0} $ & $257$ & $512$
\label{t:models}
\end{longtable}
\tablefoot{Part of the high resolution runs may not be fully resolved
  nor relaxed. We checked this has no influence on the latitudinal
  profile of the Nusselt number, but it may be of importance for its
  absolute value; hence we recommend not using it, for instance when
  studying Nusselt scalings.}

\end{document}